\documentclass{ws-rmp}
\usepackage{tikz}
\usepackage{dsfont}

\begin{document}

	\markboth{M. F. Araujo de Resende}{2D and 3D Toric Codes and the origin of their topological orders}

\catchline{}{}{}{}{}
%
%%%%%%%%%%%%%%%%%%%%%%%%%%%%%%%%%%%%%%%%%%%%%%%%%%%%%%%%%%%%%%%%%%%%

	\title{A PEDAGOGICAL OVERVIEW ON 2D AND 3D TORIC CODES AND THE ORIGIN OF THEIR TOPOLOGICAL ORDERS}

	\author{M. F. ARAUJO DE RESENDE}

	\address{Instituto de F\'{\i}sica, Universidade de S\~{a}o Paulo, 05508-090 S\~{a}o Paulo SP, Brasil \\ \email{resende@if.usp.br}}
	
	\maketitle

	\begin{history}
		\received{(03 November 2017)}
		\accepted{(17 July 2019)}
	\end{history}

	\begin{abstract}
		In this work, we will show how the topological order of the Toric Code appears when the lattice on which it is defined discretizes a three-dimensional torus. In order to do this, we will present a pedagogical review of the traditional two-dimensional Toric Code, with an emphasis on how its quasiparticles are conceived and transported. With that, we want to make clear not only how all these same quasiparticle conception and transportation fit into this three-dimensional model, but to make it clear how topology controls the degeneracy of ground state in this new situation.
	\end{abstract}

	\keywords{Toric codes; lattice gauge theories; topological order.}

	\ccode{Mathematics Subject Classification 2000: 22E46, 53C35, 57S20}

	\section{Introduction}
	
		According to the literature, many things can be said about the \emph{Toric Code} (TC), as (i) it is a toy model interpreted as a quantum error correcting code as well as it is an example of a stabilizer code \cite{kitaev-conference}, (ii) in which, although it is a self-dual model, its thermal fragility \cite{fragility} is one of the main problems for its physical realization etc. However, despite these \textquotedblleft problems for its physical realization\textquotedblright , one of the reasons why many researchers turn their attentions to this model and one of its generalizations, so-called \emph{Quantum Double Models} (QDM), is the fact that all these models have \emph{topological order} \cite{kitaev-1,pachos}.
		
		Although many experimentalists are accustomed to employing the term \textquotedblleft topological or\-der\textquotedblright \hspace*{0.01cm} when they encounter two-dimensional physical systems that, for instance, exhibit a ground state degeneracy and fractional statistics quasiparticles, we must remember that the term \textquotedblleft topological\textquotedblright \hspace*{0.01cm} is of mathematical origin \cite{pervin}. As a deduction we can affirm that, by theoretical point of view, this means that all these experimental properties are determined, in some way, by the \emph{topology} of the two-dimensional manifold where these systems are defined. By the way, this is exactly what happens with the TC and the QDM.
		
		Based on this statement, the purpose of this paper is to present a pedagogical text where, in addition to showing how the topology determines the properties mentioned above in the TC, we construct one three-dimensional generalization of this model that overcomes this two-dimensional limitation so that it is possible to evaluate the foundations of its topological order. In order to do this, we use the next Section to do a \textquotedblleft brief\textquotedblright \hspace*{0.01cm} and judicious review of the two-dimensional TC, and we use Section 3 to present the main considerations that allow us to define three-dimensional code models. Only in Section 4 we present a three-dimensional generalization of this TC by using the discretization of a three-dimensional torus and, lastly, in Section 5 we present some final remarks.
		
	\section{General properties of the Toric Code}
	
		Roughly speaking, we can say that the TC is a model whose construction is based on at least two ingredients: (i) a lattice $ \mathcal{L} _{2} $ (usually square) that discretizes a two-dimensional torus $ \mathcal{T} _{2} $; and (ii) vectors of a two-dimensional Hilbert space $ \mathfrak{H} _{2} $, which is associated with the $ j $-th edge of $ \mathcal{L} _{2} $, whose basis is $ \left\{ \left\vert g \right\rangle : g \in \mathds{Z} _{2} \right\} $. As these vectors can be interpreted as a generalization of classical bits, they are called \emph{quantum bits} (qubits) \cite{bellac}.
		
		One of the consequences of this association scheme is that, by supposing that this lattice is formed by $ N_{e} $ edges,
		\begin{equation}
			\mathfrak{H} _{\mathrm{TC}} = \mathfrak{H} ^{N_{e}} _{2} = \underbrace{ \mathfrak{H} _{2} \otimes \ldots \otimes \mathfrak{H} _{2}} _{N_{e} \ \textnormal{times}} \label{TC-hilbert}
		\end{equation}
		is the total Hilbert space associated with $ \mathcal{L} _{2} $, as long as each lattice edge corresponds to only one $ \mathfrak{H} _{2} $. In this fashion, by noting that the computational proposal behind the TC allows us to interpret it as a model that pretends to be \emph{physical}, one of its main characteristics becomes apparent: all its \textquotedblleft physical operators\textquotedblright \footnote{That is, all its operators that can measure some \textquotedblleft physical\textquotedblright \hspace*{0.01cm} property of this toy model.} that act on $ \mathfrak{H} _{\mathrm{TC}} $ must be \emph{Hermitian} \cite{mf,ballen}. One of these \textquotedblleft physical operators\textquotedblright \hspace*{0.01cm} is the \emph{Hamiltonian operator} \cite{kitaev-1}
		\begin{equation}
			H_{\mathrm{TC}} = - \sum _{v} A_{v} - \sum _{f} B_{f} \ , \label{toric-hamil}
		\end{equation}
		which is given by the linear superposition of the operators
		\begin{equation}
			A_{v} = \prod _{j \in S_{v}} \sigma ^{x} _{j} \quad \textnormal{and} \quad B_{f} = \prod _{j \in S_{f}} \sigma ^{z} _{j} \label{operators}
		\end{equation}
		such that:
		\begin{itemize}
			\item $ S_{v} $ and $ S_{f} $ are two subsets containing edges that, as shown in Figure \ref{operadores}, give structure to the $ v $-th vertex and the $ f $-th face of $ \mathcal{L} _{2} $ respectively; and
			\item the Hermitian operator \cite{kitaev-math}
			\begin{equation}
				\sigma ^{x,y,z} _{j} = \underbrace{ I_{2} \otimes \ldots \otimes I_{2} } _{\left( \textnormal{j- 1} \right) \ \textnormal{times}} \otimes \sigma ^{x,y,z} \otimes \underbrace{ I_{2} \otimes \ldots \otimes I_{2} } _{\left( N_{e} - j \right) \ \textnormal{times}} \label{sigma-composto}
			\end{equation}
			acts not identically \emph{only} on the subspace assigned to the $ j $-th edge due to its composition as a tensor product involving a single Pauli operator $ \sigma ^{x,y,z} : \mathfrak{H} _{2} \rightarrow \mathfrak{H} _{2} $ and other $ N_{e} - 1 $ identity operators.
		\end{itemize}
		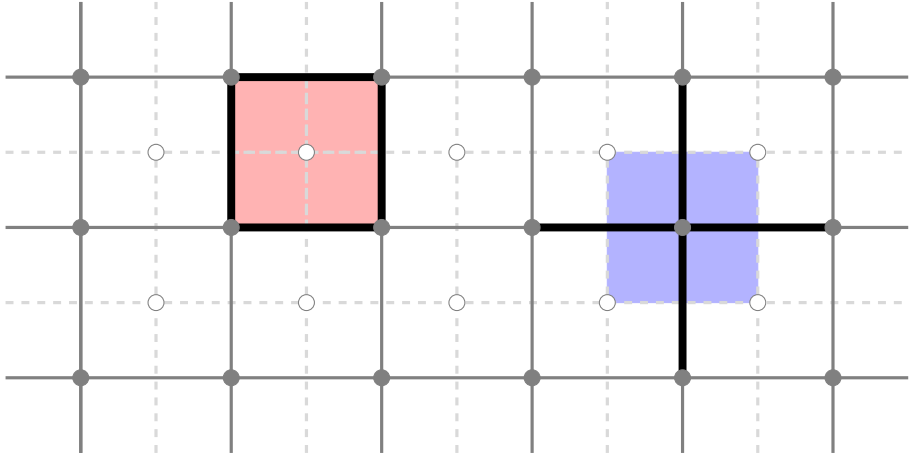
\begin{figure}[!t]
			\begin{center}
				\tikzstyle myBG=[line width=3pt,opacity=1.0]
				\newcommand{\drawLatticeLine}[2]
				{
					\draw[gray,very thick] (#1) -- (#2);
				}
				\newcommand{\drawDashedLine}[2]
				{
					\draw[gray!30,dashed,very thick] (#1) -- (#2);
				}
				\newcommand{\drawExcitedLine}[2]
				{
					\draw[black,myBG]  (#1) -- (#2);
					\draw[black,very thick] (#1) -- (#2);
				}
				\newcommand{\graphLinesHorizontal}
				{
					\drawDashedLine{3,1}{3,7};
					\drawDashedLine{5,1}{5,7};
					\drawDashedLine{7,1}{7,7};
					\drawDashedLine{9,1}{9,7};
					\drawDashedLine{11,1}{11,7};
					\drawDashedLine{1,3}{13,3};
					\drawDashedLine{1,5}{13,5};
					\drawDashedLine{4,5}{6,5};
					\drawDashedLine{5,4}{5,6};
					\drawLatticeLine{2,1}{2,7};
					\drawLatticeLine{2,1}{2,7};
					\drawLatticeLine{4,1}{4,7};
					\drawLatticeLine{6,1}{6,7};
					\drawLatticeLine{8,1}{8,7};
					\drawLatticeLine{10,1}{10,7};
					\drawLatticeLine{12,1}{12,7};
					\drawLatticeLine{1,2}{13,2};
					\drawLatticeLine{1,4}{13,4};
					\drawLatticeLine{1,6}{13,6};
					\drawExcitedLine{4,4}{4,6};
					\drawExcitedLine{6,4}{6,6};
					\drawExcitedLine{4,4}{6,4};
					\drawExcitedLine{4,6}{6,6};
					\drawExcitedLine{8,4}{12,4};
					\drawExcitedLine{10,2}{10,6};
				}
				\begin{tikzpicture}
					\draw[color=red!30,fill=red!30] (4,4) rectangle (6,6);
					\draw[color=blue!30,fill=blue!30] (9,3) rectangle (11,5);
					\graphLinesHorizontal;
					\draw[color=gray,fill=white] (3,3) circle (0.7ex);
					\draw[color=gray,fill=white] (5,3) circle (0.7ex);
					\draw[color=gray,fill=white] (7,3) circle (0.7ex);
					\draw[color=gray,fill=white] (9,3) circle (0.7ex);
					\draw[color=gray,fill=white] (11,3) circle (0.7ex);
					\draw[color=gray,fill=white] (3,5) circle (0.7ex);
					\draw[color=gray,fill=white] (5,5) circle (0.7ex);
					\draw[color=gray,fill=white] (7,5) circle (0.7ex);
					\draw[color=gray,fill=white] (9,5) circle (0.7ex);
					\draw[color=gray,fill=white] (11,5) circle (0.7ex);
					\draw[color=gray,fill=gray] (2,2) circle (0.7ex);
					\draw[color=gray,fill=gray] (4,2) circle (0.7ex);
					\draw[color=gray,fill=gray] (6,2) circle (0.7ex);
					\draw[color=gray,fill=gray] (8,2) circle (0.7ex);
					\draw[color=gray,fill=gray] (10,2) circle (0.7ex);
					\draw[color=gray,fill=gray] (12,2) circle (0.7ex);
					\draw[color=gray,fill=gray] (2,4) circle (0.7ex);
					\draw[color=gray,fill=gray] (4,4) circle (0.7ex);
					\draw[color=gray,fill=gray] (6,4) circle (0.7ex);
					\draw[color=gray,fill=gray] (8,4) circle (0.7ex);
					\draw[color=gray,fill=gray] (10,4) circle (0.7ex);
					\draw[color=gray,fill=gray] (12,4) circle (0.7ex);
					\draw[color=gray,fill=gray] (2,6) circle (0.7ex);
					\draw[color=gray,fill=gray] (4,6) circle (0.7ex);
					\draw[color=gray,fill=gray] (6,6) circle (0.7ex);
					\draw[color=gray,fill=gray] (8,6) circle (0.7ex);
					\draw[color=gray,fill=gray] (10,6) circle (0.7ex);
					\draw[color=gray,fill=gray] (12,6) circle (0.7ex);
				\end{tikzpicture}
			\end{center}
			\caption{Piece of a square lattice $ \mathcal{L} _{2} $ where we see (i) the baby blue coloured sector centred by the $ v $-th vertex of this lattice, whereas (ii) the rose-coloured sector refers to the $ f $-th face whose centroid can be interpreted as one of the vertices of a dual lattice $ \mathcal{L} ^{\ast } _{2} $ (dashed). Here, the highlighted edges (in black) define the subsets $ S_{v} $ and $ S_{f} $.}
			\label{operadores}
		\end{figure}
			
		As a matter of fact, it is good to remember that, when we say that an operator is a Hamiltonian, this operator must be able to measure the total \emph{energy} of the system it represents. And the best way to understand how $ H_{\mathrm{TC}} $ measures the TC energy is by analysing the behaviour of operators (\ref{operators}) that define it, more specifically when these operators act on one of the $ 2^{N_{e}} $ elements
		\begin{equation}
			\left\vert \varphi _{1} \right\rangle \otimes \ldots \otimes \left\vert \varphi _{N_{e}} \right\rangle \label{elemento}
		\end{equation}
		of a basis of $ \mathfrak{H} _{\mathrm{TC}} $. After all, since Hilbert spaces are examples of vector spaces, if we understand this action we will understand what appears from the action of these operators on any element of $ \mathfrak{H} _{\mathrm{TC}} $ \cite{elon}. By the way, as (\ref{elemento}) is a tensor product of the $ N_{e} $ elements $ \left\vert 0 \right\rangle $ and $ \left\vert 1 \right\rangle $ \footnote{Elements which can be interpreted as the eigenvectors of the Pauli operator $ \sigma ^{z} $ in a matrix representation \cite{arfken}
		\begin{equation}
			\sigma ^{x}	=
			\begin{pmatrix}
				\ 0 & 1 \ \\
				\ 1 & 0 \
			\end{pmatrix} \ , \quad \sigma ^{y}	=
			\begin{pmatrix}
				\ 0 & -i \ \\
				\ i & 0
			\end{pmatrix} \quad \textnormal{and} \quad \sigma ^{z} =
			\begin{pmatrix}
				\ 1 & 0 \\
				\ 0 & -1 \
			\end{pmatrix} \label{pauli}
		\end{equation}
		that rests on a basis $ \left\{ \left\vert 0 \right\rangle , \left\vert 1 \right\rangle \right\} $ where \label{base-escolhida-page}
		\begin{equation*}
			\left\vert 0 \right\rangle =
			\begin{pmatrix}
				1 \\
				\ 0 \
			\end{pmatrix} \quad \textnormal{and} \quad \left\vert 1 \right\rangle =
			\begin{pmatrix}
				0 \\
				\ 1 \
			\end{pmatrix} \ .
		\end{equation*}},
		an interesting conclusion arises: the vertex ($ A_{v} $) and face ($ B_{f} $) operators defined in (\ref{operators}) \emph{commute}.
		
		Although the main result that help us to understand the behaviour of the Hamiltonian (\ref{toric-hamil}) as an \textquotedblleft energy meter\textquotedblright \hspace*{0.01cm} is
		\begin{equation*}
			\left( A_{v} \right) ^{2} = \left( B_{f} \right) ^{2} = \underbrace{ I_{2} \otimes \ldots \otimes I_{2}} _{N_{e} \ \textnormal{times}} = \mathds{1} _{v,p}
		\end{equation*}
		(because it shows us that the eigenvalues of $ A_{v} $ and $ B_{f} $ are only equal to $ 1 $ and $ - 1 $), the commutativity among these vertex and face operators has great relevance for this understanding. After all, as
		\begin{itemize}
			\item the measurement performed by $ A_{v} $ does not interfere in the measurement performed by $ B_{f} $ and vice-versa, and
			\item the smallest energy $ E_{0} = - \left( N_{v} + N_{f} \right) $ associated with this system occurs when
			\begin{equation}
				A_{v} \bigl\vert \xi _{0} \bigr\rangle = \bigl\vert \xi _{0} \bigr\rangle \quad \textnormal{and} \quad B_{f} \bigl\vert \xi _{0} \bigr\rangle = \bigl\vert \xi _{0} \bigr\rangle \label{protege}
			\end{equation}
			is satisfied for all $ N_{v} $ vertices and $ N_{f} $ faces that structure $ \mathcal{L} _{2} $,
		\end{itemize}
		this commutative property allows to conclude that \cite{cirac}
		\begin{equation}
			\bigl\vert \xi ^{\left( 1 \right) } _{0} \bigr\rangle = \frac{1}{\sqrt{2}} \prod _{v} \left( \mathds{1} _{v} + A_{v} \right) \underbrace{ \left\vert 0 \right\rangle \otimes \ldots \otimes \left\vert 0 \right\rangle } _{N_{e} \textnormal{times}} \label{vacuo-1}
		\end{equation}
		is an eigenstate that satisfies (\ref{protege}); in other words, (\ref{vacuo-1}) is one of the TC \emph{vacuum states}.
			
		\subsection{About the ground state}
		
			Evidently, there is a good reason we say that (\ref{vacuo-1}) is just \emph{one} of the vacuum states: \emph{there are others}. Nevertheless, before we present these other vacuum states, we think it is more pedagogical to show to the reader how it is possible to withdraw the system from any of its vacuum states. After all, as this will allow us to understand gradually how the entire TC energy spectrum is built, it will make us understand how the topology controls the non-univocity of the ground state.
			
			\subsubsection{First excitations}
				
				In order to understand the TC energy spectrum, let us consider a state
				\begin{equation}
					\left\vert \xi ^{\prime } \right\rangle = \sigma ^{x} _{k} \bigl\vert \xi ^{ \left( 1 \right) } _{0} \bigr\rangle \ . \label{nao-vacuo-exemplo-1}
				\end{equation}
				By considering that $ \alpha $ and $ \beta $ are any of the super indexes $ x $, $ y $ and $ z $, as the commuting properties between Pauli operators $ \sigma ^{\alpha } $ and $ \sigma ^{\beta } $ lead us to
				\begin{equation}
					\sigma ^{\alpha } _{j} \circ \sigma ^{\beta } _{k} =
					\begin{cases}
						\ - \sigma ^{\beta } _{k} \circ \sigma ^{\alpha } _{j} , \ \textnormal{if} \ \alpha \neq \beta \ \textnormal{and} \ j = k , \ \textnormal{and} \\
						\hspace*{0.4cm} \sigma ^{\beta } _{k} \circ \sigma ^{\alpha } _{j} , \ \textnormal{otherwise} ,
					\end{cases} \label{sigma-comutacao}
				\end{equation}
				this allows to note that
				\begin{equation}
					B_{f^{\prime }} \left\vert \xi ^{\prime } \right\rangle = \left( \prod _{j \in S_{f^{\prime }}} \sigma ^{z} _{j} \right) \circ \sigma ^{x} _{k} \left\vert \xi ^{ \left( 1 \right) } _{0} \right\rangle = - \sigma ^{x} _{k} \circ \left( \prod _{j \in S_{f^{\prime }}} \sigma ^{z} _{j} \left\vert \xi ^{ \left( 1 \right) } _{0} \right\rangle \right) = - \sigma ^{x} _{k} \left\vert \xi ^{ \left( 1 \right) } _{0} \right\rangle = - \left\vert \xi ^{\prime } \right\rangle \label{b-neg-1}
				\end{equation}
				when the $ k $-th lattice edge belongs to $ S_{f^{\prime }} $; that is, $ \bigl\vert \xi ^{\prime } \bigr\rangle $ does not satisfy \emph{at least} one of the conditions listed in (\ref{protege}) and, therefore, cannot be considered as a vacuum state. Furthermore, as $ \mathcal{L} _{2} $ discretizes the surface of a two-dimensional torus and, then, (\ref{b-neg-1}) also applies to a $ f^{\prime \prime } $-th face such that $ k = S_{f^{\prime }} \cap S_{f^{\prime \prime }} $, the energy related to non-vacuum (\ref{nao-vacuo-exemplo-1}) is
				\begin{equation*}
					E_{1} = E_{0} + 4
				\end{equation*}
				because, for $ N_{v} $ vertices and other $ N_{f} - 2 $ faces that complete $ \mathcal{L} _{2} $, we have
				\begin{equation*}
					A_{v} \left\vert \xi ^{\prime } \right\rangle = \left\vert \xi ^{\prime } \right\rangle \quad \textnormal{and} \quad B_{f} \left\vert \xi ^{\prime } \right\rangle = \left\vert \xi ^{\prime } \right\rangle \ .
				\end{equation*}
					
				An entirely analogous comment also applies to state
				\begin{equation}
					\left\vert \xi ^{\prime \prime } \right\rangle = \sigma ^{z} _{k} \bigl\vert \xi ^{ \left( 1 \right) } _{0} \bigr\rangle \label{nao-vacuo-exemplo-2}
				\end{equation}
				where the $ k $-th lattice edge belongs to $ S_{v^{\prime }} \cap S_{v^{\prime \prime }} $. In this new case, as (\ref{sigma-comutacao}) implies not only that
				\begin{equation*}
					A_{v ^{\prime }} \left\vert \xi ^{\prime \prime } \right\rangle = A_{v ^{\prime \prime }} \left\vert \xi ^{\prime \prime } \right\rangle = - \left\vert \xi ^{\prime \prime } \right\rangle \ ,
				\end{equation*}
				but also ensures that
				\begin{equation*}
					A_{v} \left\vert \xi ^{\prime \prime } \right\rangle = \left\vert \xi ^{\prime \prime } \right\rangle \quad \textnormal{and} \quad B_{f} \left\vert \xi ^{\prime \prime } \right\rangle = \left\vert \xi ^{\prime \prime } \right\rangle
				\end{equation*}
				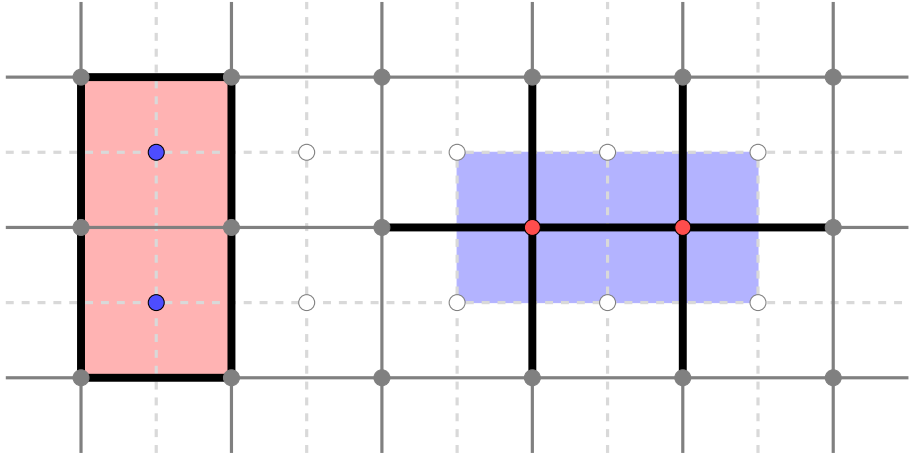
\begin{figure}[!t]
					\begin{center}
						\tikzstyle myBG=[line width=3pt,opacity=1.0]
						\newcommand{\drawLatticeLine}[2]
						{
							\draw[gray,very thick] (#1) -- (#2);
						}
						\newcommand{\drawDashedLine}[2]
						{
							\draw[gray!30,dashed,very thick] (#1) -- (#2);
						}
						\newcommand{\drawExcitationLine}[2]
						{
							\draw[black,myBG]  (#1) -- (#2);
							\draw[black,very thick] (#1) -- (#2);
						}
						\newcommand{\graphLinesHorizontal}
						{
							\drawDashedLine{3,1}{3,7};
							\drawDashedLine{5,1}{5,7};
							\drawDashedLine{7,1}{7,7};
							\drawDashedLine{9,1}{9,7};
							\drawDashedLine{11,1}{11,7};
							\drawDashedLine{1,3}{13,3};
							\drawDashedLine{1,5}{13,5};
							\drawLatticeLine{2,1}{2,7};
							\drawLatticeLine{4,1}{4,7};
							\drawLatticeLine{6,1}{6,7};
							\drawLatticeLine{8,1}{8,7};
							\drawLatticeLine{10,1}{10,7};
							\drawLatticeLine{12,1}{12,7};
							\drawLatticeLine{1,2}{13,2};
							\drawLatticeLine{1,4}{13,4};
							\drawLatticeLine{1,6}{13,6};
							\drawExcitationLine{2,2}{2,6};
							\drawExcitationLine{4,2}{4,6};
							\drawExcitationLine{2,2}{4,2};
							\drawExcitationLine{2,6}{4,6};
							\drawExcitationLine{6,4}{12,4};
							\drawExcitationLine{8,2}{8,6};
							\drawExcitationLine{10,2}{10,6};
						}
						\begin{tikzpicture}
							\draw[color=red!30,fill=red!30] (2,2) rectangle (4,6);
							\draw[color=blue!30,fill=blue!30] (7,3) rectangle (11,5);
							\graphLinesHorizontal;
							\draw[color=black,fill=blue!70] (3,3) circle (0.7ex);
							\draw[color=gray,fill=white] (5,3) circle (0.7ex);
							\draw[color=gray,fill=white] (7,3) circle (0.7ex);
							\draw[color=gray,fill=white] (9,3) circle (0.7ex);
							\draw[color=gray,fill=white] (11,3) circle (0.7ex);
							\draw[color=black,fill=blue!70] (3,5) circle (0.7ex);
							\draw[color=gray,fill=white] (5,5) circle (0.7ex);
							\draw[color=gray,fill=white] (7,5) circle (0.7ex);
							\draw[color=gray,fill=white] (9,5) circle (0.7ex);
							\draw[color=gray,fill=white] (11,5) circle (0.7ex);
							\draw[color=gray,fill=gray] (2,2) circle (0.7ex);
							\draw[color=gray,fill=gray] (4,2) circle (0.7ex);
							\draw[color=gray,fill=gray] (6,2) circle (0.7ex);
							\draw[color=gray,fill=gray] (8,2) circle (0.7ex);
							\draw[color=gray,fill=gray] (10,2) circle (0.7ex);
							\draw[color=gray,fill=gray] (12,2) circle (0.7ex);
							\draw[color=gray,fill=gray] (2,4) circle (0.7ex);
							\draw[color=gray,fill=gray] (4,4) circle (0.7ex);
							\draw[color=gray,fill=gray] (6,4) circle (0.7ex);
							\draw[color=black,fill=red!70] (8,4) circle (0.7ex);
							\draw[color=black,fill=red!70] (10,4) circle (0.7ex);
							\draw[color=gray,fill=gray] (12,4) circle (0.7ex);
							\draw[color=gray,fill=gray] (2,6) circle (0.7ex);
							\draw[color=gray,fill=gray] (4,6) circle (0.7ex);
							\draw[color=gray,fill=gray] (6,6) circle (0.7ex);
							\draw[color=gray,fill=gray] (8,6) circle (0.7ex);
							\draw[color=gray,fill=gray] (10,6) circle (0.7ex);
							\draw[color=gray,fill=gray] (12,6) circle (0.7ex);
						\end{tikzpicture}
					\end{center}
					\caption{Piece of $ \mathcal{L} _{2} $ where: two quasiparticles (in red) were created, in two neighbouring vertices (baby blue coloured sector), due to the action of the operator $ \sigma ^{z} _{k} $; and two quasiparticles (in blue) were created, in two neighbouring faces (rose-coloured sector), due to the action of the operator $ \sigma ^{x} _{j} $.}
					\label{cargas-m}
				\end{figure}
				are still valid for the remaining $ N_{v} - 2 $ vertices and $ N_{f} $ faces that complete $ \mathcal{L} _{2} $, we can conclude that the energy of (\ref{nao-vacuo-exemplo-2}) also is $ E_{1} $ because the same number of constraints that defines the ground state was again violated.
				
				According to all that we have just presented, it is immediate to conclude that these vertex and face operators are unable to change any of (\ref{vacuo-1}), (\ref{nao-vacuo-exemplo-1}) or (\ref{nao-vacuo-exemplo-2}): the only thing that these operators actually do is check whether these states can be identified \emph{locally} as a vacuum or not, further reinforcing the counting scheme performed by the Hamiltonian (\ref{toric-hamil}). Anyhow, due to the capacity that $ A_{v} $ and $ B_{f} $ have to localize excitations on the vertices and faces of $ \mathcal{L} _{2} $, we need to clarify two statements.
				
				The first one is that, as illustrated in Figure \ref{cargas-m}, all these local elevations of energy can be realized as \emph{quasiparticles}:
				\begin{itemize}
					\item in the case related to eigenstate (\ref{nao-vacuo-exemplo-1}), for instance, we can associate this non-vacuum (which is signalled only on two neighbouring faces $ f^{\prime } $ and $ f^{\prime \prime } $) to the existence of two quasiparticles (one on each of these faces);
					\item in the case of the eigenstate (\ref{nao-vacuo-exemplo-2}), this quasiparticle association is due to the signalization of non-vacuum only on two neighbouring vertices $ v^{\prime } $ and $ v^{\prime \prime } $ (scilicet, one excitation on each vertex).
				\end{itemize}
				
				The second statement that we must clarify here is a consequence of this quasiparticle realization. After all, although the quasiparticles associated to (\ref{nao-vacuo-exemplo-1}) and (\ref{nao-vacuo-exemplo-2}) have the same energy, we cannot prove they are the same because they are detected by operators which act effectively in different lattice sectors. In spite of this distinction argument is perfectly correct, it may be weaker than another we will give soon. However, as these quasiparticles may really be different from each other, from now on we will denote those that are detectable by the vertex operators as \emph{type e}, whereas those that are detectable by the face operators are denoted as \emph{type m}.
				
			\subsubsection{The transport of quasiparticles}
				
				Due to what we have seen so far, it is not hard to conclude that (\ref{nao-vacuo-exemplo-1}) and (\ref{nao-vacuo-exemplo-2}) are not the unique non-vacuum states of the TC, nor is it difficult to conclude that the energies associated with all these possible non-vacuums can assume only values $ E_{n} = E_{0} + 4n $, where $ n $ is a non-null natural number. Yet, in order to understand the properties of the quasiparticles associated with these possible non-vacuums, instead of \textquotedblleft going crazy\textquotedblright \hspace*{0.01cm} trying to list all them, we can make a simpler and more general analysis by using states such as
				\begin{equation}
					\left\vert \xi \right\rangle = \sigma ^{\alpha } _{k} \bigl\vert \xi ^{ \left( 1 \right) } _{0} \bigr\rangle \ . \label{hypothetical-state}
				\end{equation}
				And the property that will be most useful for this analysis is
				\begin{equation}
					\bigl( \sigma ^{\alpha } _{j} \bigr) ^{2} = \mathds{1} _{j} \ . \label{id-prop}
				\end{equation}
				
				By the way, something that is clear of (\ref{id-prop}) is that, as the action of $ \sigma ^{\alpha } _{k} $ on the $ \left\vert \xi \right\rangle $ causes the system to return to the vacuum condition (\ref{vacuo-1}), \emph{all} quasiparticles created by this operator can be interpreted as their own \emph{anti-quasiparticles}: when an operator
				\begin{equation}
					O^{ \alpha } _{jk} = \sigma ^{\alpha } _{j} \circ \sigma ^{\alpha } _{k} \label{TC-teletrans-operator}
				\end{equation}
				acts on (\ref{hypothetical-state}), it annihilates a quasiparticle pair in the vicinity of a $ k $-th edge at the same time that it will create another (with the same properties as the one who was annihilated) in the adjacencies of the $ j $-th edge. In this fashion, one of the more natural interpretations that emerges is that (\ref{TC-teletrans-operator}) can be seen as a kind of \textquotedblleft quasiparticle pair teleport operator\textquotedblright ; strictly speaking, it is an operator that causes a quasiparticle pair, which was in the vicinity of a $ k $-th edge, to reappear around a $ j $-th edge that may be completely arbitrary.
				
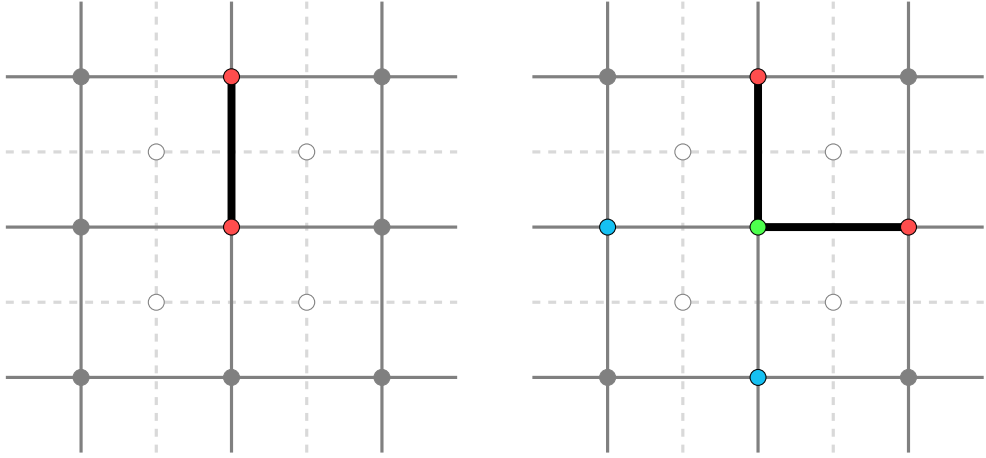
\begin{figure}[!t]
					\begin{center}
						\tikzstyle myBG=[line width=3pt,opacity=1.0]
						\newcommand{\drawLatticeLine}[2]
						{
							\draw[gray,very thick] (#1) -- (#2);
						}
						\newcommand{\drawDashedLine}[2]
						{
							\draw[gray!30,dashed,very thick] (#1) -- (#2);
						}
						\newcommand{\drawExcitationLine}[2]
						{
							\draw[black,thin,myBG]  (#1) -- (#2);
							\draw[black,very thick] (#1) -- (#2);
						}
						\newcommand{\graphLinesHorizontal}
						{
							\drawDashedLine{3,1}{3,7};
							\drawDashedLine{5,1}{5,7};
							\drawDashedLine{10,1}{10,7};
							\drawDashedLine{12,1}{12,7};
							\drawDashedLine{1,3}{7,3};
							\drawDashedLine{8,3}{14,3};
							\drawDashedLine{1,5}{7,5};
							\drawDashedLine{8,5}{14,5};
							\drawLatticeLine{2,1}{2,7};
							\drawLatticeLine{4,1}{4,7};
							\drawLatticeLine{6,1}{6,7};
							\drawLatticeLine{9,1}{9,7};
							\drawLatticeLine{11,1}{11,7};
							\drawLatticeLine{13,1}{13,7};
							\drawLatticeLine{1,2}{7,2};
							\drawLatticeLine{8,2}{14,2};
							\drawLatticeLine{1,4}{7,4};
							\drawLatticeLine{8,4}{14,4};
							\drawLatticeLine{1,6}{7,6};
							\drawLatticeLine{8,6}{14,6};
							\drawExcitationLine{4,4}{4,6};
							\drawExcitationLine{11,4}{11,6};
							\drawExcitationLine{11,4}{13,4};
						}
						\begin{tikzpicture}
							\graphLinesHorizontal;
							\draw[color=gray,fill=white] (3,3) circle (0.7ex);
							\draw[color=gray,fill=white] (5,3) circle (0.7ex);
							\draw[color=gray,fill=white] (10,3) circle (0.7ex);
							\draw[color=gray,fill=white] (12,3) circle (0.7ex);
							\draw[color=gray,fill=white] (3,5) circle (0.7ex);
							\draw[color=gray,fill=white] (5,5) circle (0.7ex);
							\draw[color=gray,fill=white] (10,5) circle (0.7ex);
							\draw[color=gray,fill=white] (12,5) circle (0.7ex);
							\draw[color=gray,fill=gray] (2,2) circle (0.7ex);
							\draw[color=gray,fill=gray] (4,2) circle (0.7ex);
							\draw[color=gray,fill=gray] (6,2) circle (0.7ex);
							\draw[color=gray,fill=gray] (9,2) circle (0.7ex);
							\draw[color=black,fill=cyan!70] (11,2) circle (0.7ex);
							\draw[color=gray,fill=gray] (13,2) circle (0.7ex);
							\draw[color=gray,fill=gray] (2,4) circle (0.7ex);
							\draw[color=black,fill=red!70] (4,4) circle (0.7ex);
							\draw[color=gray,fill=gray] (6,4) circle (0.7ex);
							\draw[color=black,fill=cyan!70] (9,4) circle (0.7ex);
							\draw[color=black,fill=green!70] (11,4) circle (0.7ex);
							\draw[color=black,fill=red!70] (13,4) circle (0.7ex);
							\draw[color=gray,fill=gray] (2,6) circle (0.7ex);
							\draw[color=black,fill=red!70] (4,6) circle (0.7ex);
							\draw[color=gray,fill=gray] (6,6) circle (0.7ex);
							\draw[color=gray,fill=gray] (9,6) circle (0.7ex);
							\draw[color=black,fill=red!70] (11,6) circle (0.7ex);
							\draw[color=gray,fill=gray] (13,6) circle (0.7ex);
						\end{tikzpicture}
					\end{center}
					\caption{On left we have two quasiparticles $ e $ (in red, in two vertices of $ \mathcal{L} _{2} $) due to the action of a single operator $ \sigma ^{z} _{k} $ where before it was a vacuum. On right we have same lattice region in the posterior time, where only one of these quasiparticles was annihilated by the action of $ \sigma ^{z} _{j} $ because $ j $ and $ k $ are indexing adjacent edges. In this last case, as such annihilation was accompanied by the creation of a new quasiparticle over a third vertex, we can interpret this process as a quasiparticle transport over $ \mathcal{L} _{2} $. Note that the other two vertices (highlighted in cyan, which also belong to the unit circle centred in the green point) correspond to the other two positions to which this quasiparticle $ e $ could have been carried by an operator $ \sigma ^{z} _{j^{\prime }} $.}
					\label{transporte-z}
				\end{figure}

				Although the term \textquotedblleft teleportation\textquotedblright \hspace*{0.01cm} always raises interesting ideas, even more interesting is the what arises from the action of a single $ O^{z} _{jk} $ on (\ref{hypothetical-state}) when $ j $ and $ k $ are indexing two distinct edges that belong to the same $ S_{v} $. After all, since (i) these two lattice edges share a same vertex $ v $ and (ii) the superposition of two quasiparticles $ e $ at any vertex is identified as a local vacuum, the number of quasiparticles associated with the system after the action of $ O^{z} _{jk} $ remains intact. Thus, if we analyse the action of this operator in the \textquotedblleft disassembled way\textquotedblright \hspace*{0.01cm} that, for instance, is shown in Figure \ref{transporte-z},
				it is possible to interpret the disappearance of the quasiparticle $ e $ (which was in the $ v $-th vertex) as its transport to one of the neighbouring vertices which belong to unit circle in a \emph{Taxicab Geometry} \cite{bianconi}. Moreover, by the same reasoning (which leads to $ O^{z} _{jk} $) also leads us to a more general transport operator
				\begin{equation}
					O^{z} _{\boldsymbol{\gamma }} = \prod _{j \in \boldsymbol{\gamma }} \sigma ^{z} _{j} \label{caminho-z}
				\end{equation}
				where $ \boldsymbol{\gamma } $ is a \emph{path} consisting of a greater number of edges two to two neighbours, when we consider $ \boldsymbol{\gamma } $ as \emph{closed} another interesting result becomes evident: as for this closed path there is no quasiparticle associated,
				\begin{equation}
					\bigl\vert \xi ^{\left( 1 \right) } _{0} \bigr\rangle ^{\prime } = O^{z} _{\boldsymbol{\gamma }} \bigl\vert \xi ^{\left( 1 \right) } _{0} \bigr\rangle \label{vacuo-2}
				\end{equation}				
				is also a vacuum state of the TC.
				
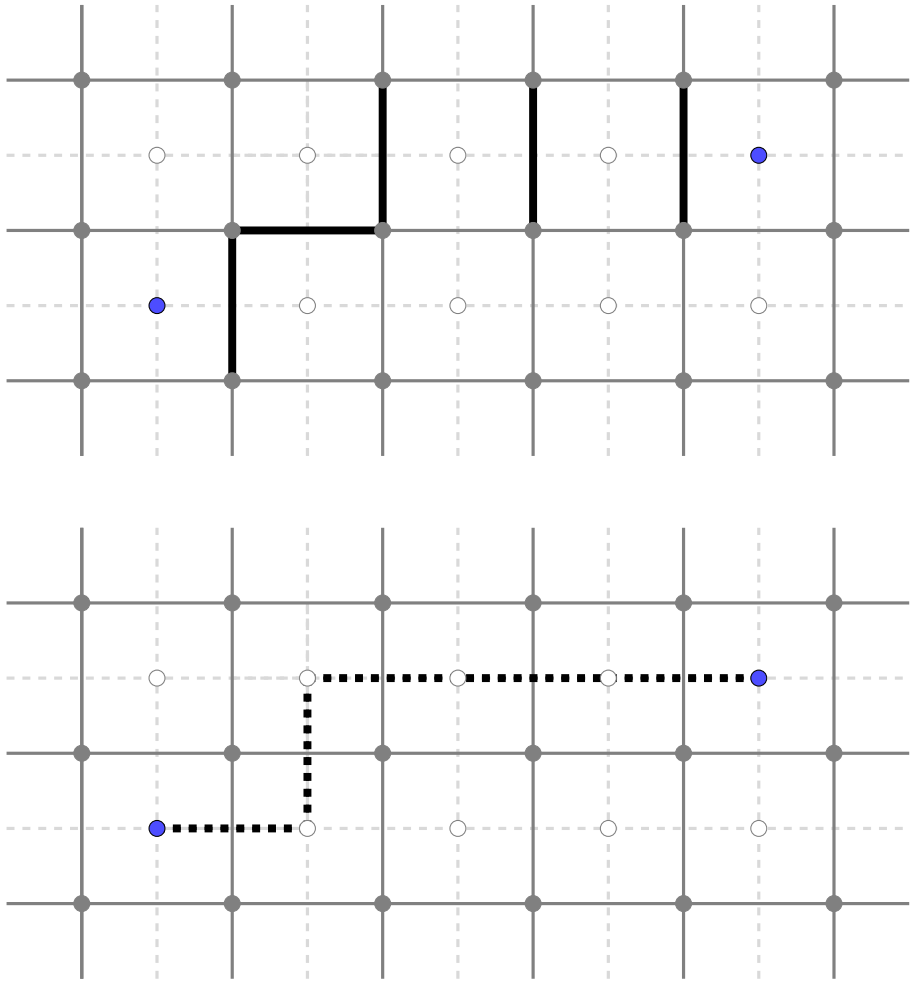
\begin{figure}[!t]
					\begin{center}
						\tikzstyle myBG=[line width=3pt,opacity=1.0]
						\newcommand{\drawLatticeLine}[2]
						{
							\draw[gray,very thick] (#1) -- (#2);
						}
						\newcommand{\drawDashedLine}[2]
						{
							\draw[gray!30,dashed,very thick] (#1) -- (#2);
						}
						\newcommand{\drawExcitedLine}[2]
						{
							\draw[black,myBG]  (#1) -- (#2);
							\draw[black,very thick] (#1) -- (#2);
						}
						\newcommand{\graphLinesHorizontal}
						{
							\drawDashedLine{3,1}{3,7};
							\drawDashedLine{5,1}{5,7};
							\drawDashedLine{7,1}{7,7};
							\drawDashedLine{9,1}{9,7};
							\drawDashedLine{11,1}{11,7};
							\drawDashedLine{1,3}{13,3};
							\drawDashedLine{1,5}{13,5};
							\drawDashedLine{4,5}{6,5};
							\drawDashedLine{5,4}{5,6};
							\drawLatticeLine{2,1}{2,7};
							\drawLatticeLine{2,1}{2,7};
							\drawLatticeLine{4,1}{4,7};
							\drawLatticeLine{6,1}{6,7};
							\drawLatticeLine{8,1}{8,7};
							\drawLatticeLine{10,1}{10,7};
							\drawLatticeLine{12,1}{12,7};
							\drawLatticeLine{1,2}{13,2};
							\drawLatticeLine{1,4}{13,4};
							\drawLatticeLine{1,6}{13,6};
							\drawExcitedLine{4,2}{4,4};
							\drawExcitedLine{6,4}{6,6};
							\drawExcitedLine{4,4}{6,4};
							\drawExcitedLine{8,4}{8,6};
							\drawExcitedLine{10,4}{10,6};
						}
						\begin{tikzpicture}
							\graphLinesHorizontal;
							\draw[color=black,fill=blue!70] (3,3) circle (0.7ex);
							\draw[color=gray,fill=white] (5,3) circle (0.7ex);
							\draw[color=gray,fill=white] (7,3) circle (0.7ex);
							\draw[color=gray,fill=white] (9,3) circle (0.7ex);
							\draw[color=gray,fill=white] (11,3) circle (0.7ex);
							\draw[color=gray,fill=white] (3,5) circle (0.7ex);
							\draw[color=gray,fill=white] (5,5) circle (0.7ex);
							\draw[color=gray,fill=white] (7,5) circle (0.7ex);
							\draw[color=gray,fill=white] (9,5) circle (0.7ex);
							\draw[color=black,fill=blue!70] (11,5) circle (0.7ex);
							\draw[color=gray,fill=gray] (2,2) circle (0.7ex);
							\draw[color=gray,fill=gray] (4,2) circle (0.7ex);
							\draw[color=gray,fill=gray] (6,2) circle (0.7ex);
							\draw[color=gray,fill=gray] (8,2) circle (0.7ex);
							\draw[color=gray,fill=gray] (10,2) circle (0.7ex);
							\draw[color=gray,fill=gray] (12,2) circle (0.7ex);
							\draw[color=gray,fill=gray] (2,4) circle (0.7ex);
							\draw[color=gray,fill=gray] (4,4) circle (0.7ex);
							\draw[color=gray,fill=gray] (6,4) circle (0.7ex);
							\draw[color=gray,fill=gray] (8,4) circle (0.7ex);
							\draw[color=gray,fill=gray] (10,4) circle (0.7ex);
							\draw[color=gray,fill=gray] (12,4) circle (0.7ex);
							\draw[color=gray,fill=gray] (2,6) circle (0.7ex);
							\draw[color=gray,fill=gray] (4,6) circle (0.7ex);
							\draw[color=gray,fill=gray] (6,6) circle (0.7ex);
							\draw[color=gray,fill=gray] (8,6) circle (0.7ex);
							\draw[color=gray,fill=gray] (10,6) circle (0.7ex);
							\draw[color=gray,fill=gray] (12,6) circle (0.7ex);
						\end{tikzpicture}
					\end{center} \bigskip
					\begin{center}
						\tikzstyle myBG=[line width=3pt,opacity=1.0]
						\newcommand{\drawLatticeLine}[2]
						{
							\draw[gray,very thick] (#1) -- (#2);
						}
						\newcommand{\drawDashedLine}[2]
						{
							\draw[gray!30,dashed,very thick] (#1) -- (#2);
						}
						\newcommand{\drawExcitedLine}[2]
						{
							\draw[black,myBG]  (#1) -- (#2);
							\draw[black,very thick] (#1) -- (#2);
						}
						\newcommand{\drawDashedExcitedLine}[2]
						{
							\draw[black,dashed,myBG]  (#1) -- (#2);
							\draw[black,dashed,very thick] (#1) -- (#2);
						}
						\newcommand{\graphLinesHorizontal}
						{
							\drawDashedLine{3,1}{3,7};
							\drawDashedLine{5,1}{5,7};
							\drawDashedLine{7,1}{7,7};
							\drawDashedLine{9,1}{9,7};
							\drawDashedLine{11,1}{11,7};
							\drawDashedLine{1,3}{13,3};
							\drawDashedLine{1,5}{13,5};
							\drawDashedLine{4,5}{6,5};
							\drawDashedLine{5,4}{5,6};
							\drawLatticeLine{2,1}{2,7};
							\drawLatticeLine{2,1}{2,7};
							\drawLatticeLine{4,1}{4,7};
							\drawLatticeLine{6,1}{6,7};
							\drawLatticeLine{8,1}{8,7};
							\drawLatticeLine{10,1}{10,7};
							\drawLatticeLine{12,1}{12,7};
							\drawLatticeLine{1,2}{13,2};
							\drawLatticeLine{1,4}{13,4};
							\drawLatticeLine{1,6}{13,6};
							\drawDashedExcitedLine{3,3}{5,3};
							\drawDashedExcitedLine{5,3}{5,5};
							\drawDashedExcitedLine{5,5}{11,5};
						}
						\begin{tikzpicture}
							\graphLinesHorizontal;
							\draw[color=black,fill=blue!70] (3,3) circle (0.7ex);
							\draw[color=gray,fill=white] (5,3) circle (0.7ex);
							\draw[color=gray,fill=white] (7,3) circle (0.7ex);
							\draw[color=gray,fill=white] (9,3) circle (0.7ex);
							\draw[color=gray,fill=white] (11,3) circle (0.7ex);
							\draw[color=gray,fill=white] (3,5) circle (0.7ex);
							\draw[color=gray,fill=white] (5,5) circle (0.7ex);
							\draw[color=gray,fill=white] (7,5) circle (0.7ex);
							\draw[color=gray,fill=white] (9,5) circle (0.7ex);
							\draw[color=black,fill=blue!70] (11,5) circle (0.7ex);
							\draw[color=gray,fill=gray] (2,2) circle (0.7ex);
							\draw[color=gray,fill=gray] (4,2) circle (0.7ex);
							\draw[color=gray,fill=gray] (6,2) circle (0.7ex);
							\draw[color=gray,fill=gray] (8,2) circle (0.7ex);
							\draw[color=gray,fill=gray] (10,2) circle (0.7ex);
							\draw[color=gray,fill=gray] (12,2) circle (0.7ex);
							\draw[color=gray,fill=gray] (2,4) circle (0.7ex);
							\draw[color=gray,fill=gray] (4,4) circle (0.7ex);
							\draw[color=gray,fill=gray] (6,4) circle (0.7ex);
							\draw[color=gray,fill=gray] (8,4) circle (0.7ex);
							\draw[color=gray,fill=gray] (10,4) circle (0.7ex);
							\draw[color=gray,fill=gray] (12,4) circle (0.7ex);
							\draw[color=gray,fill=gray] (2,6) circle (0.7ex);
							\draw[color=gray,fill=gray] (4,6) circle (0.7ex);
							\draw[color=gray,fill=gray] (6,6) circle (0.7ex);
							\draw[color=gray,fill=gray] (8,6) circle (0.7ex);
							\draw[color=gray,fill=gray] (10,6) circle (0.7ex);
							\draw[color=gray,fill=gray] (12,6) circle (0.7ex);
						\end{tikzpicture}
					\end{center}
					\caption{In the figure above, one pair of quasiparticles $ m $ (in blue) was created due to the action of operators $ \sigma ^{x} _{j} $ on the subspace associated with the highlighted lattice edges (in black). In the figure below, we have the same situation seen from the point of view of the dual lattice $ \mathcal{L} ^{\ast } _{2} $: that is, this same quasiparticle pair was created by action of the same operators $ \sigma ^{x} _{j} $ on the dual lattice edges (highlighted in dashed black) that are two to two neighbours.}
					\label{caminho-dual}
				\end{figure}

				Obviously this same transport interpretation also extends to an operator
				\begin{equation}
					O^{x} _{\boldsymbol{\gamma } ^{\ast }} = \prod _{j \in \boldsymbol{\gamma } ^{\ast }} \sigma ^{x} _{j} \label{caminho-x}
				\end{equation}
				that uses a path $ \boldsymbol{\gamma } ^{\ast } $ composed of edges that are two to two neighbours by the point of view of the dual lattice $ \mathcal{L} ^{\ast } _{2} $, such as shown in Figure \ref{caminho-dual},
				as well as this vacuum interpretation also extends, for instance, to the other eigenstates
				\begin{equation}
					\bigl\vert \xi ^{\left( 1 \right) } _{0} \bigr\rangle ^{\prime \prime } = O^{x} _{\boldsymbol{\gamma } ^{\ast }} \bigl\vert \xi ^{\left( 1 \right) } _{0} \bigr\rangle \quad \textnormal{and} \quad \bigl\vert \xi ^{\left( 1 \right) } _{0} \bigr\rangle ^{\prime \prime \prime } = O^{x} _{\boldsymbol{\gamma } ^{\ast }} \circ O^{z} _{\boldsymbol{\gamma }} \bigl\vert \xi ^{\left( 1 \right) } _{0} \bigr\rangle \label{vacuo-3}
				\end{equation}
				if $ \boldsymbol{\gamma } ^{\ast } $ is also a closed path. However, it is precisely in the light of these (\ref{vacuo-1}), (\ref{vacuo-2}) and (\ref{vacuo-3}) vacuum states that we need to make an important observation, which is specifically related to the fact that the TC is defined on a two-dimensional torus discretization.
				
			\subsubsection{The degeneracy of the ground state}
			
				When we take a lattice such as $ \mathcal{L} _{2} $, it is not difficult to realize the large number of closed paths $ \boldsymbol{\gamma } $ and $ \boldsymbol{\gamma } ^{\ast } $ that can be defined by using its edges, provided that this lattice discretizes a manifold with a good resolution. Yet, as $ \mathcal{L} _{2} $ discretizes a two-dimensional torus, it is important to note that some of these closed paths have a specific property: they are \emph{non-contractile}.
				
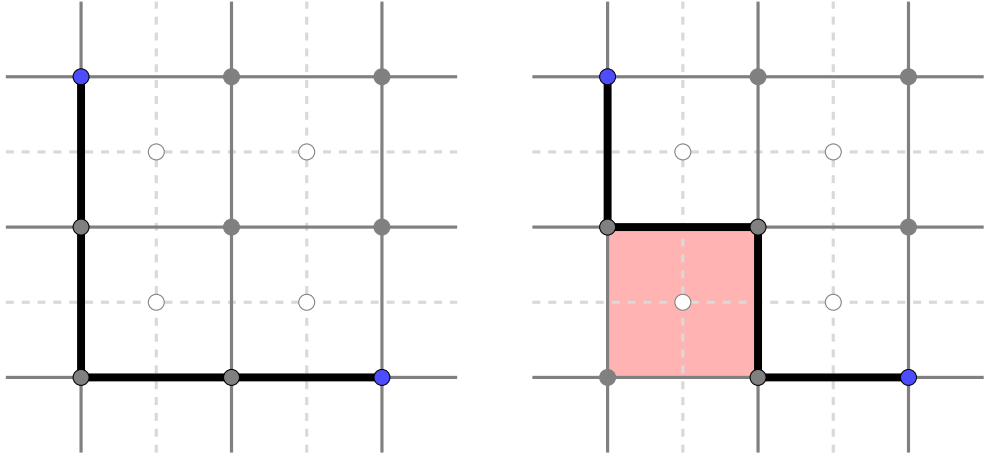
\begin{figure}[!t]
					\begin{center}
						\tikzstyle myBG=[line width=3pt,opacity=1.0]
						\newcommand{\drawLatticeLine}[2]
						{
							\draw[gray,very thick] (#1) -- (#2);
						}
						\newcommand{\drawDashedLine}[2]
						{
							\draw[gray!30,dashed,very thick] (#1) -- (#2);
						}
						\newcommand{\drawExcitedLine}[2]
						{
							\draw[black,myBG]  (#1) -- (#2);
							\draw[black,very thick] (#1) -- (#2);
						}
						\newcommand{\graphLinesHorizontal}
						{
							\drawDashedLine{3,1}{3,7};
							\drawDashedLine{5,1}{5,7};
							\drawDashedLine{10,1}{10,7};
							\drawDashedLine{12,1}{12,7};
							\drawDashedLine{1,3}{7,3};
							\drawDashedLine{8,3}{14,3};
							\drawDashedLine{1,5}{7,5};
							\drawDashedLine{8,5}{14,5};
							\drawLatticeLine{2,1}{2,7};
							\drawLatticeLine{4,1}{4,7};
							\drawLatticeLine{6,1}{6,7};
							\drawLatticeLine{9,1}{9,7};
							\drawLatticeLine{11,1}{11,7};
							\drawLatticeLine{13,1}{13,7};
							\drawLatticeLine{1,2}{7,2};
							\drawLatticeLine{8,2}{14,2};
							\drawLatticeLine{1,4}{7,4};
							\drawLatticeLine{8,4}{14,4};
							\drawLatticeLine{1,6}{7,6};
							\drawLatticeLine{8,6}{14,6};
							\drawExcitedLine{2,2}{6,2};
							\drawExcitedLine{11,2}{13,2};
							\drawExcitedLine{9,4}{11,4};
							\drawExcitedLine{2,2}{2,6};
							\drawExcitedLine{9,4}{9,6};
							\drawExcitedLine{11,2}{11,4};
						}
						\begin{tikzpicture}
							\draw[color=red!30,fill=red!30] (9,2) rectangle (11,4);
							\graphLinesHorizontal;
							\draw[color=gray,fill=white] (3,3) circle (0.7ex);
							\draw[color=gray,fill=white] (5,3) circle (0.7ex);
							\draw[color=gray,fill=white] (10,3) circle (0.7ex);
							\draw[color=gray,fill=white] (12,3) circle (0.7ex);
							\draw[color=gray,fill=white] (3,5) circle (0.7ex);
							\draw[color=gray,fill=white] (5,5) circle (0.7ex);
							\draw[color=gray,fill=white] (10,5) circle (0.7ex);
							\draw[color=gray,fill=white] (12,5) circle (0.7ex);
							\draw[color=black,fill=gray] (2,2) circle (0.7ex);
							\draw[color=black,fill=gray] (4,2) circle (0.7ex);
							\draw[color=black,fill=blue!70] (6,2) circle (0.7ex);
							\draw[color=gray,fill=gray] (9,2) circle (0.7ex);
							\draw[color=black,fill=gray] (11,2) circle (0.7ex);
							\draw[color=black,fill=blue!70] (13,2) circle (0.7ex);
							\draw[color=black,fill=gray] (2,4) circle (0.7ex);
							\draw[color=gray,fill=gray] (4,4) circle (0.7ex);
							\draw[color=gray,fill=gray] (6,4) circle (0.7ex);
							\draw[color=black,fill=gray] (9,4) circle (0.7ex);
							\draw[color=black,fill=gray] (11,4) circle (0.7ex);
							\draw[color=gray,fill=gray] (13,4) circle (0.7ex);
							\draw[color=black,fill=blue!70] (2,6) circle (0.7ex);
							\draw[color=gray,fill=gray] (4,6) circle (0.7ex);
							\draw[color=gray,fill=gray] (6,6) circle (0.7ex);
							\draw[color=black,fill=blue!70] (9,6) circle (0.7ex);
							\draw[color=gray,fill=gray] (11,6) circle (0.7ex);
							\draw[color=gray,fill=gray] (13,6) circle (0.7ex);
						\end{tikzpicture}
					\end{center}
					\caption{On left we have a pair of quasiparticles $ e $ that was created by the action of an operator $ O^{z} _{\boldsymbol{\gamma }} $ on the path $ \boldsymbol{\gamma } $ of the edges highlighted (in black). On right we see the deformation of this same path after the action of a single $ B_{p} $ on one of the faces (rose-coloured sector) that share edges with $ \boldsymbol{\gamma } $ due to property (\ref{id-prop}).}
					\label{deformation}
				\end{figure}

				Despite the concept of \emph{contraction} is better understood for closed curves in a manifold (provided there are continuous applications which shrinking these closed curves to a single point \cite{hatcher}), the characterization of a contractile path on $ \mathcal{L} _{2} $ is due to the possibility of reducing any (\ref{vacuo-2}) or (\ref{vacuo-3}) to the first vacuum state (\ref{vacuo-1}). In order to understand how all this works in the TC, it is interesting to note that, as
				\begin{itemize}
					\item the set $ S_{f} $ is composed by all the edges that belong to the boundary of the $ f $-th face of $ \mathcal{L} _{2} $, whereas
					\item each of the edges of $ S_{v} $ intersects only one of the dual edges that complete the $ v $-th face of $ \mathcal{L} ^{\ast } _{2} $ (or, equivalently, the $ v $-th dual face of $ \mathcal{L} $),
				\end{itemize}
				$ A_{v} $ and $ B_{f} $ cannot transport a quasiparticle to a position other than the one it occupies: as illustrated in Figure \ref{deformation},
				due to (\ref{id-prop}) these operators can only deform the (closed) paths in which they act. In this way, when we consider that $ F_{A} $ and $ F_{B} $ are two applications composed by a finite number of vertex and face operators respectively, since
				\begin{equation}
					F_{A} \circ O^{x} _{\boldsymbol{\gamma } ^{\ast }} = O^{x} _{\boldsymbol{\gamma } ^{\ast }} \circ F_{A} = \mathds{1} _{\mathrm{TC}} \quad \textnormal{and} \quad F_{B} \circ O^{z} _{\boldsymbol{\gamma }} = O^{z} _{\boldsymbol{\gamma }} \circ F_{B} = \mathds{1} _{\mathrm{TC}} \label{contract-condition}
				\end{equation}
				are valid only if $ \boldsymbol{\gamma } ^{\ast } $ and $ \boldsymbol{\gamma } $ correspond to the discretization of contractile paths whose interiors contain the vertices and faces on which $ F_{A} $ and $ F_{B} $ act respectively, we conclude that the non-contractility of some path (whether dual or not) is defined via the non-verification of some equality in (\ref{contract-condition}).
				
\begin{figure}[!t]
					\centering
					\includegraphics[viewport=280 25 0 130,scale=1.5]{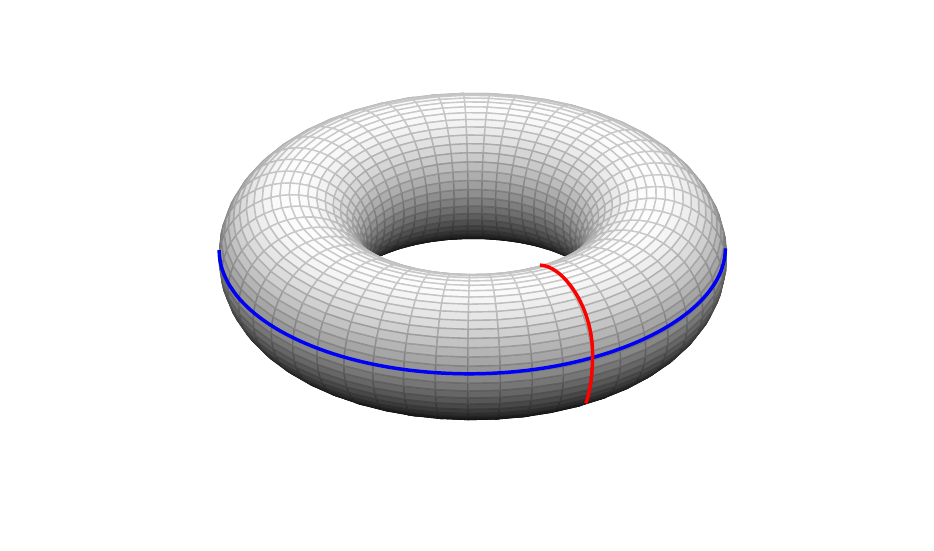}
					\caption{Here we have a torus $ \mathcal{T} _{2} $ discretized by the square lattice $ \mathcal{L} _{2} $. Note that, although it is possible to define contractile paths in $ \mathcal{L} _{2} $, two curves deserve attention: one $ \bar{\boldsymbol{\gamma }} _{1} $ (in red) that contour the $ \mathcal{T} _{2} $ loop, and another $ \bar{\boldsymbol{\gamma }} _{2} $ (in blue) that contour the hole that characterizes $ \mathcal{T} _{2} $ as a torus with genus one. These two closed paths, as well as all their possible deformations, are examples of non-contractile paths in $ \mathcal{L} _{2} $, which should be interpreted as the discretization of the curves that generate the \emph{fundamental group} of $ \mathcal{T} _{2} $ (i.e., the \emph{first homotopy group} $ \pi _{1} \left( \mathcal{T} _{2} \right) $).}
					\label{torus}
				\end{figure}

				Indeed, the simple fact that $ \mathcal{L} _{2} $ discretizes a two-dimensional torus implies the existence of non-contractile closed paths $ \bar{\boldsymbol{\gamma }} $ and $ \bar{\boldsymbol{\gamma }} ^{\ast } $, as shown in Figure \ref{torus}. And the main consequence of this is that, unlike vacuum states (\ref{vacuo-2}) and (\ref{vacuo-3}) are reducible to the first (\ref{vacuo-1}), the states
				\begin{equation*}
					\bigl\vert \xi ^{\left( 2 \right) } _{0} \bigr\rangle = O^{ \alpha } _{\bar{\mathcal{C}} _{1}} \bigl\vert \xi ^{\left( 1 \right) } _{0} \bigr\rangle \ , \quad \bigl\vert \xi ^{\left( 3 \right) } _{0} \bigr\rangle = O^{ \beta } _{\bar{\mathcal{C}} _{2}} \bigl\vert \xi ^{\left( 1 \right) } _{0} \bigr\rangle \quad \textnormal{and} \quad \bigl\vert \xi ^{\left( 4 \right) } _{0} \bigr\rangle = O^{ \alpha } _{\bar{\mathcal{C}} _{1}} \circ O^{ \beta } _{\bar{\mathcal{C}} _{2}} \bigl\vert \xi ^{\left( 1 \right) } _{0} \bigr\rangle \ ,
				\end{equation*}
				which are defined by using curves that are non-contractile in distinct directions, do not \emph{seem} to be reducible to neither (\ref{vacuo-1}) or to each other. Here, the symbol $ \bar{\mathcal{C}} $ should be interpreted as either: the non-contractile $ \bar{\boldsymbol{\gamma }} $, if the indices $ \alpha $ and $ \beta $ are equal to $ z $; or the non-contractile $ \bar{\boldsymbol{\gamma }} ^{\ast } $, if these indices are equal to $ x $.
				
				In spite of this observation of non-reducibility makes perfect sense, we need to elucidate why we said \textquotedblleft seem\textquotedblright \hspace*{0.01cm} in the last paragraph. And the reason we use this term is very simple: as $ \sigma ^{z} $ is completely unable to make any change
				\begin{equation}
					\bigl\vert 0 \bigr\rangle \leftrightarrow \bigl\vert 1 \bigr\rangle \ , \label{troca-TC}
				\end{equation}
				$ O^{z} _{\bar{\boldsymbol{\gamma }}} $ is also completely unable to change the encoding that defines any TC state. In other words, despite all the non-contractility of $ \bar{\boldsymbol{\gamma }} $, an eigenstate such as
				\begin{equation}
					\bigl\vert \xi ^{\left( 2 \right) } _{0} \bigr\rangle = O^{z} _{\bar{\boldsymbol{\gamma }}} \bigl\vert \xi ^{\left( 1 \right) } _{0} \bigr\rangle \label{non-new-vacuum}
				\end{equation}
				does not define a new eigenstate that is fundamentally independent of (\ref{vacuo-1}): only
				\begin{equation}
					\bigl\vert \xi ^{\left( 2 \right) } _{0} \bigr\rangle = O^{x} _{\bar{\boldsymbol{\gamma }} ^{\ast } _{1}} \bigl\vert \xi ^{\left( 1 \right) } _{0} \bigr\rangle \ , \quad \bigl\vert \xi ^{\left( 3 \right) } _{0} \bigr\rangle = O^{x} _{\bar{\boldsymbol{\gamma }} ^{\ast } _{2}} \bigl\vert \xi ^{\left( 1 \right) } _{0} \bigr\rangle \quad \textnormal{and} \quad \bigl\vert \xi ^{\left( 4 \right) } _{0} \bigr\rangle = O^{x} _{\bar{\boldsymbol{\gamma }} ^{\ast } _{1}} \circ O^{x} _{\bar{\boldsymbol{\gamma }} ^{\ast } _{2}} \bigl\vert \xi ^{\left( 1 \right) } _{0} \bigr\rangle \label{vacuo-degenerado}
				\end{equation}
				correspond to vacuum states independent of each other and with respect to the first (\ref{vacuo-1}) because their constructions rest on exchanges (\ref{troca-TC}). Another way of understanding this point is by noting that every face operator is a local \textquotedblleft holonomy meter\textquotedblright : thus, as the fact that $ B_{f} $ and $ \sigma ^{z} $ commute can be interpreted as the inability of $ \sigma ^{z} $ to deform the manifold that $ \mathcal{L} _{2} $ discretizes, (\ref{non-new-vacuum}) really does not define a new vacuum state by the topological point of view.
				
				As the matter of fact, it is this \emph{four-fold degeneracy} of the ground state that makes clear one part of the topological aspect behind this model. After all, when we analyse (\ref{vacuo-degenerado}) by using a physical point of view, all this non-contractility of closed paths can be associated with four physical quantities that are \emph{conserved}, i.e., four physical quantities which cannot be exterminated by the action of vertex or/and face operators. By the way, as these four conserved quantities correspond to four different topological sectors, whose existence can only be detected by the action of the operators
				\begin{equation*}
					O^{z} _{\bar{\boldsymbol{\gamma }} _{1}} \ \textnormal{and} \ O^{z} _{\bar{\boldsymbol{\gamma }} _{2}} \ , 
				\end{equation*}
				each of these sectors ends up being connected to one of the vacuum states in (\ref{vacuo-1}) and (\ref{vacuo-degenerado}). Therefore, if we consider a situation of very low temperatures, for example, we can give a physical interpretation to this four-fold degeneracy by associating it with four different phases (coexisting in the same temperature regime) which exist only due to the topology of the manifold where this system is defined \cite{castelnovo}, what reinforces the comment made at the end of the last paragraph.
				
		\subsection{A new quasiparticle}
		
			Just to complete all comments that may be related to the creation and transport of quasiparticles, it is worth mentioning that, albeit we have already presented $ \sigma ^{x} _{j} $ and $ \sigma ^{z} _{j} $ as active agents in all these processes, nothing has been said about $ \sigma ^{y} _{j} $, whose definition has already been given in (\ref{sigma-composto}).
			
			In order to try to redeem ourselves and justify all this apparent omission, it is essential to begin by noting that, if we consider
			\begin{equation}
				\left\vert \xi ^{\prime \prime \prime } \right\rangle = \sigma ^{y} _{a} \bigl\vert \xi ^{ \left( 1 \right) } _{0} \bigr\rangle \label{nao-vacuo-exemplo-3}
			\end{equation}	
			as a third example of non-vacuum state, the relations (\ref{sigma-comutacao}) lead us to
			\begin{eqnarray*}
				A_{v} \left\vert \xi ^{\prime \prime \prime } \right\rangle = - \left\vert \xi ^{\prime \prime \prime } \right\rangle & , & A_{v ^{\prime }} \left\vert \xi ^{\prime \prime \prime } \right\rangle = - \left\vert \xi ^{\prime \prime \prime } \right\rangle \ , \\
				B_{f} \left\vert \xi ^{\prime \prime \prime } \right\rangle = - \left\vert \xi ^{\prime \prime \prime } \right\rangle & \ \textnormal{and} \ & B_{p^{\prime }} \left\vert \xi ^{\prime \prime \prime } \right\rangle = - \left\vert \xi ^{\prime \prime \prime } \right\rangle \ ,
			\end{eqnarray*}
			where $ a = S_{v} \cap S_{v ^{\prime }} \cap S_{f} \cap S_{p^{\prime }} $ is indexing the single edge on which $ \sigma ^{y} _{a} $ acts effectively. Scilicet, $ \sigma ^{y} _{j} $ creates alone four excitations: one in each of the vertices that enclose this edge, one in each of the faces that share this edge, which raise the energy of this lattice system to
			\begin{equation*}
				E_{2} = E_{0} + 8 \ .
			\end{equation*}	
			Although (\ref{nao-vacuo-exemplo-3}) seems to behave as a non-vacuum state, there are at least two peculiarities related to this joint excitation that deserves our full attention, and the first of them refers to its transport.
				
			\subsubsection{First peculiarity}
				
				In an attempt to extend the transport mechanism presented above to the excitations created by $ \sigma ^{y} _{j} $, it is quite natural to take an operator
				\begin{equation}
					O^{y} _{ba} = \sigma ^{y} _{b} \circ \sigma ^{y} _{a} \ . \label{TC-teletrans-y}
				\end{equation}
				After all, as the action of this operator on (\ref{nao-vacuo-exemplo-3}) would annihilate one quadruple excitation of the $ a $-th edge adjacencies and would create another, with the same properties, around the $ b $-th edge, it would be very comfortable if the only purpose was to keep the system with the same energy. However, if we remember that the possibility of transport excitations by operators which do not identify themselves as a \textquotedblleft teleportation\textquotedblright \hspace*{0.01cm} (\ref{TC-teletrans-y}) is real, it is also natural to make this transport using operators such as
				\begin{equation}
					O^{y} _{\boldsymbol{\gamma }} = \prod _{j \in \boldsymbol{\gamma }} \sigma ^{y} _{j} \quad \textnormal{or/and} \quad O^{y} _{\boldsymbol{\gamma } ^{\ast }} = \prod _{j \in \boldsymbol{\gamma } ^{\ast }} \sigma ^{y} _{j} \ . \label{caminho-y}
				\end{equation}
				
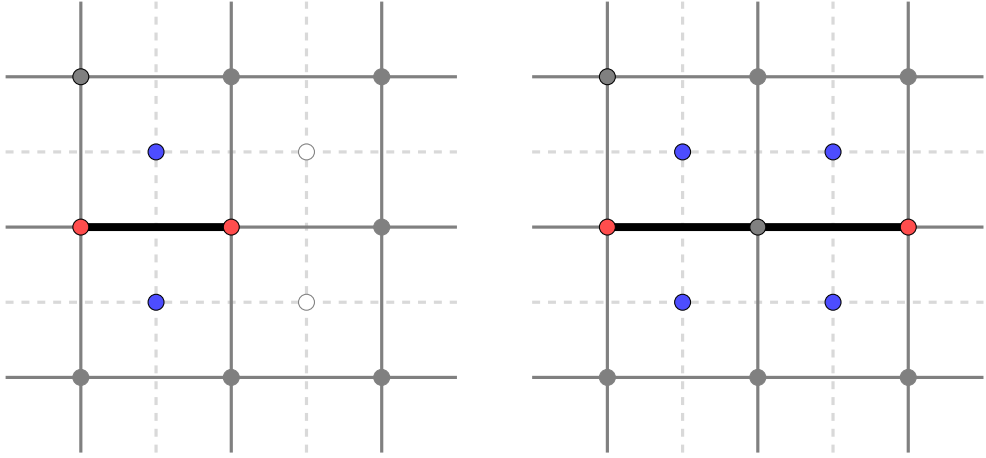
\begin{figure}[!t]
					\begin{center}
						\tikzstyle myBG=[line width=3pt,opacity=1.0]
						\newcommand{\drawLatticeLine}[2]
						{
							\draw[gray,very thick] (#1) -- (#2);
						}
						\newcommand{\drawDashedLine}[2]
						{
							\draw[gray!30,dashed,very thick] (#1) -- (#2);
						}
						\newcommand{\drawExcitedLine}[2]
						{
							\draw[black,myBG]  (#1) -- (#2);
							\draw[black,very thick] (#1) -- (#2);
						}
						\newcommand{\graphLinesHorizontal}
						{
							\drawDashedLine{3,1}{3,7};
							\drawDashedLine{5,1}{5,7};
							\drawDashedLine{10,1}{10,7};
							\drawDashedLine{12,1}{12,7};
							\drawDashedLine{1,3}{7,3};
							\drawDashedLine{8,3}{14,3};
							\drawDashedLine{1,5}{7,5};
							\drawDashedLine{8,5}{14,5};
							\drawLatticeLine{2,1}{2,7};
							\drawLatticeLine{4,1}{4,7};
							\drawLatticeLine{6,1}{6,7};
							\drawLatticeLine{9,1}{9,7};
							\drawLatticeLine{11,1}{11,7};
							\drawLatticeLine{13,1}{13,7};
							\drawLatticeLine{1,2}{7,2};
							\drawLatticeLine{8,2}{14,2};
							\drawLatticeLine{1,4}{7,4};
							\drawLatticeLine{8,4}{14,4};
							\drawLatticeLine{1,6}{7,6};
							\drawLatticeLine{8,6}{14,6};
							\drawExcitedLine{2,4}{4,4};
							\drawExcitedLine{9,4}{13,4};
						}
						\begin{tikzpicture}
							\graphLinesHorizontal;
							\draw[color=black,fill=blue!70] (3,3) circle (0.7ex);
							\draw[color=gray,fill=white] (5,3) circle (0.7ex);
							\draw[color=black,fill=blue!70] (10,3) circle (0.7ex);
							\draw[color=black,fill=blue!70] (12,3) circle (0.7ex);
							\draw[color=black,fill=blue!70] (3,5) circle (0.7ex);
							\draw[color=gray,fill=white] (5,5) circle (0.7ex);
							\draw[color=black,fill=blue!70] (10,5) circle (0.7ex);
							\draw[color=black,fill=blue!70] (12,5) circle (0.7ex);
							\draw[color=gray,fill=gray] (2,2) circle (0.7ex);
							\draw[color=gray,fill=gray] (4,2) circle (0.7ex);
							\draw[color=gray,fill=gray] (6,2) circle (0.7ex);
							\draw[color=gray,fill=gray] (9,2) circle (0.7ex);
							\draw[color=gray,fill=gray] (11,2) circle (0.7ex);
							\draw[color=gray,fill=gray] (13,2) circle (0.7ex);
							\draw[color=black,fill=red!70] (2,4) circle (0.7ex);
							\draw[color=black,fill=red!70] (4,4) circle (0.7ex);
							\draw[color=gray,fill=gray] (6,4) circle (0.7ex);
							\draw[color=black,fill=red!70] (9,4) circle (0.7ex);
							\draw[color=black,fill=gray] (11,4) circle (0.7ex);
							\draw[color=black,fill=red!70] (13,4) circle (0.7ex);
							\draw[color=black,fill=gray] (2,6) circle (0.7ex);
							\draw[color=gray,fill=gray] (4,6) circle (0.7ex);
							\draw[color=gray,fill=gray] (6,6) circle (0.7ex);
							\draw[color=black,fill=gray] (9,6) circle (0.7ex);
							\draw[color=gray,fill=gray] (11,6) circle (0.7ex);
							\draw[color=gray,fill=gray] (13,6) circle (0.7ex);
						\end{tikzpicture}
					\end{center}
					\caption{On left we see a single compound excitation, which is generated by the action of a single $ \sigma ^{z} _{a} $. On right we see exactly the same lattice region where, after the action of the operator $ \sigma ^{z} _{b} $ on one of the neighbouring edges of $ a $, a single vertex excitation was transported. Note that, in this transportation, the total energy of the system is not conserved.}
					\label{transporte-y}
				\end{figure}

				By considering this latter possibility, it is perfectly reasonable to consider $ O^{y} _{\boldsymbol{\gamma }} $ as the simplest transport operators, where $ \boldsymbol{\gamma } $ is a path formed only by the $ b $-th edge such that $ \left\{ a , b \right\} \in S_{v} $. Just as it is not difficult to observe that the number of excitations detectable by vertex operators will be \emph{conserved} along this transport, it is also not difficult to note that the same thing cannot necessarily be said about the excitations detectable by face operators. After all, as shown in Figure \ref{transporte-y},
				if these two edges are such that
				\begin{equation*}
					a \in S_{f} \cap S_{p^{\prime }} \quad \textnormal{and} \quad \left( b \notin S_{f} \right) \vee \left( b \notin S_{p^{\prime }} \right) \ ,
				\end{equation*}
				the number of excitations detectable by the face operators will \emph{increase}: this increasing of energy does not happen only when
				\begin{equation}
					a \in S_{f} \cap S_{p^{\prime }} \quad \textnormal{and} \quad b \in S_{f} \cup S_{p^{\prime }} \cup S_{v} \cup S_{v ^{\prime }} \ . \label{y-condition}
				\end{equation}
				In this fashion, it is clear that, in the more general case where $ \boldsymbol{\gamma } $ is structured by a greater number of edges two to two adjacent, the transport of these excitations without any addition of energy is possible only if (\ref{y-condition}) is satisfied for each pair of edges that composes $ \boldsymbol{\gamma } $.
				
			\subsubsection{Second peculiarity}
			
				Although an analogous conclusion also follows for the action of a $ O^{y} _{\boldsymbol{\gamma } ^{\ast }} $ on (\ref{nao-vacuo-exemplo-3}), it is interesting to explore the second peculiarity which we have suggested to exist. And to understand this second peculiarity, we need to turn our attention to the non-vacuum states
				\begin{equation}
					\left\vert \xi ^{\prime \prime \prime \prime } \right\rangle = \sigma ^{x} _{j} \circ \sigma ^{y} _{j} \bigl\vert \xi ^{ \left( 1 \right) } _{0} \bigr\rangle \quad \textnormal{and} \quad \left\vert \xi ^{\prime \prime \prime \prime \prime } \right\rangle = \sigma ^{z} _{j} \circ \sigma ^{y} _{j} \bigl\vert \xi ^{ \left( 1 \right) } _{0} \bigr\rangle \ .\label{nao-vacuo-exemplo-4}
				\end{equation}
				After all, as (\ref{sigma-comutacao}) also imply that
				\begin{equation*}
					A_{v} \left\vert \xi ^{\prime \prime \prime \prime } \right\rangle = - \left\vert \xi ^{\prime \prime \prime \prime } \right\rangle \quad \textnormal{and} \quad B_{f} \left\vert \xi ^{\prime \prime \prime \prime } \right\rangle = \left\vert \xi ^{\prime \prime \prime \prime } \right\rangle \ ,
				\end{equation*}
				it becomes clear that, if this $ j $-th edge belongs to $ S_{v} \cap S_{f} $, the operator $ \sigma ^{x} _{j} \circ \sigma ^{y} _{j} $ can be effectively interpreted as the same $ \sigma ^{z} _{j} $ that is capable of creating only quasiparticles $ e $.
				
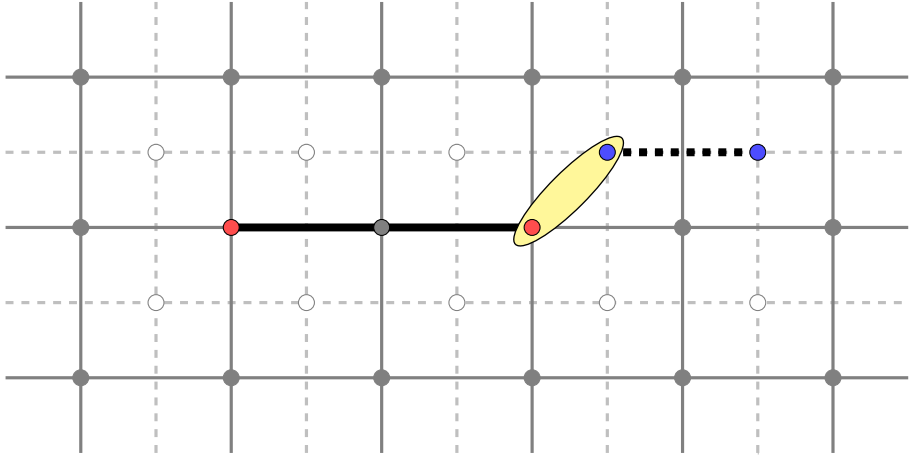
\begin{figure}[!t]
					\begin{center}
						\tikzstyle myBG=[line width=3pt,opacity=1.0]
						\newcommand{\drawLatticeLine}[2]
						{
							\draw[gray,very thick] (#1) -- (#2);
						}
						\newcommand{\drawDashedLine}[2]
						{
							\draw[gray!50,dashed,very thick] (#1) -- (#2);
						}
						\newcommand{\drawExcitedLine}[2]
						{
							\draw[black,myBG]  (#1) -- (#2);
							\draw[black,very thick] (#1) -- (#2);
						}
						\newcommand{\drawExcitedDashedLine}[2]
						{
							\draw[black,dashed,myBG]  (#1) -- (#2);
							\draw[black,dashed,very thick] (#1) -- (#2);
						}
						\newcommand{\graphLinesHorizontal}
						{
							\drawDashedLine{3,1}{3,7};
							\drawDashedLine{5,1}{5,7};
							\drawDashedLine{7,1}{7,7};
							\drawDashedLine{9,1}{9,7};
							\drawDashedLine{11,1}{11,7};
							\drawDashedLine{11,1}{11,1};
							\drawDashedLine{1,3}{13,3};
							\drawDashedLine{1,5}{13,5};
							\drawLatticeLine{2,1}{2,7};
							\drawLatticeLine{4,1}{4,7};
							\drawLatticeLine{6,1}{6,7};
							\drawLatticeLine{8,1}{8,7};
							\drawLatticeLine{10,1}{10,7};
							\drawLatticeLine{12,1}{12,7};
							\drawLatticeLine{1,2}{13,2};
							\drawLatticeLine{1,4}{13,4};
							\drawLatticeLine{1,6}{13,6};
							\drawExcitedLine{4,4}{8,4};
							\drawExcitedDashedLine{9,5}{11,5};
						}
						\begin{tikzpicture}
							\graphLinesHorizontal;
							\draw [rotate=45,fill=yellow!50,thin,line width=0.5pt] (8.17,-2.83) arc[x radius=1.0cm, y radius =0.25cm, start angle=-180, end angle=180];
							\draw[color=gray,fill=white] (3,3) circle (0.7ex);
							\draw[color=gray,fill=white] (5,3) circle (0.7ex);
							\draw[color=gray,fill=white] (7,3) circle (0.7ex);
							\draw[color=gray,fill=white] (9,3) circle (0.7ex);
							\draw[color=gray,fill=white] (11,3) circle (0.7ex);
							\draw[color=gray,fill=white] (3,5) circle (0.7ex);
							\draw[color=gray,fill=white] (5,5) circle (0.7ex);
							\draw[color=gray,fill=white] (7,5) circle (0.7ex);
							\draw[color=black,fill=blue!70] (9,5) circle (0.7ex);
							\draw[color=black,fill=blue!70] (11,5) circle (0.7ex);
							\draw[color=gray,fill=gray] (2,2) circle (0.7ex);
							\draw[color=gray,fill=gray] (4,2) circle (0.7ex);
							\draw[color=gray,fill=gray] (6,2) circle (0.7ex);
							\draw[color=gray,fill=gray] (8,2) circle (0.7ex);
							\draw[color=gray,fill=gray] (10,2) circle (0.7ex);
							\draw[color=gray,fill=gray] (12,2) circle (0.7ex);
							\draw[color=gray,fill=gray] (2,4) circle (0.7ex);
							\draw[color=black,fill=red!70] (4,4) circle (0.7ex);
							\draw[color=black,fill=gray] (6,4) circle (0.7ex);
							\draw[color=black,fill=red!70] (8,4) circle (0.7ex);
							\draw[color=gray,fill=gray] (10,4) circle (0.7ex);
							\draw[color=gray,fill=gray] (12,4) circle (0.7ex);
							\draw[color=gray,fill=gray] (2,6) circle (0.7ex);
							\draw[color=gray,fill=gray] (4,6) circle (0.7ex);
							\draw[color=gray,fill=gray] (6,6) circle (0.7ex);
							\draw[color=gray,fill=gray] (8,6) circle (0.7ex);
							\draw[color=gray,fill=gray] (10,6) circle (0.7ex);
							\draw[color=gray,fill=gray] (12,6) circle (0.7ex);
						\end{tikzpicture}
					\end{center}
					\caption{Schematic drawing where we can identify one dyon (in yellow) as a pair of quasiparticles (one type $ e $ and another type $ m $) in the vicinity of a single edge.}
					\label{dyon}
				\end{figure}

				Analogously, since (\ref{nao-vacuo-exemplo-4}) also allows to observe that
				\begin{equation*}
					A_{v} \left\vert \xi ^{\prime \prime \prime \prime \prime } \right\rangle = \left\vert \xi ^{\prime \prime \prime \prime \prime } \right\rangle \quad \textnormal{and} \quad B_{f} \left\vert \xi ^{\prime \prime \prime \prime \prime } \right\rangle = - \left\vert \xi ^{\prime \prime \prime \prime \prime } \right\rangle \ ,
				\end{equation*}
				$ \sigma ^{z} _{j} \circ \sigma ^{y} _{j} $ can also be effectively seen as $ \sigma ^{x} _{j} $ because both can create only quasiparticles with properties similar to those of type $ m $. In this way, in view of
				\begin{itemize}
					\item the fact that this new operator $ \sigma ^{y} _{j} $ could create a pair of independent excitations of the previous ones, and
					\item the desire to construct the TC as a model whose excitations can be transportable without any addition of energy,
				\end{itemize}
				this effectivity leads us to the definition of an operator
				\begin{equation}
					\textnormal{\textquotedblleft } \sigma ^{y } _{j} \textnormal{\textquotedblright } = \sigma ^{x} _{j} \circ \sigma ^{z} _{j} = \sigma ^{z} _{j} \circ \sigma ^{x} _{j} \ , \label{producao-epsilon}
				\end{equation}
				that has nothing new. The only novelty here is that, as $ \mathds{1} _{j} $ and $ \textnormal{\textquotedblleft } \sigma ^{y } _{j} \textnormal{\textquotedblright } $ are the two elements that allow us to identify
				\begin{equation}
					\left\{ \mathds{1} _{j} , \sigma ^{x} _{j} , \textnormal{\textquotedblleft } \sigma ^{y } _{j} \textnormal{\textquotedblright } , \sigma ^{z} _{j} \right\} \label{TC-abelian}
				\end{equation}
				as an \emph{Abelian group}, $ \textnormal{\textquotedblleft } \sigma ^{y} _{j} \textnormal{\textquotedblright } $ should be interpreted as the creator of a new excitation defined as the union of a quasiparticle $ e $ with another of type $ m $ in the adjacencies of a single edge, as shown in Figure \ref{dyon}. This new quasiparticle, which we will label as type $ \epsilon $, it is recognized as a \emph{dyon}.
					
			\subsubsection{A small parenthesis}
			
				Anyhow, it is important to take the opportunity provided by (\ref{producao-epsilon}) to look at what looks like the third peculiarity related to this joint excitation that we have just defined as a dyon: this $ \epsilon $ is created via the same operators that define the Hamiltonian (\ref{toric-hamil}). In this fashion, by noting that all the quasiparticles $ e $ and $ m $ are created by these same operators, we arrive at the following conclusion: as well as in QFT, where Hamiltonians can be expressed in the \emph{Fock representation} by using the creation $ a^{\dagger } $ and annihilation  $ a $ operators \cite{itzykson}, the entire TC energy spectrum can also be well understood from
				\begin{itemize}
					\item the knowledge of the ground state of this model, and
					\item the excitations created by the action of the operators that compose its Hamiltonian (\ref{toric-hamil}) on this ground state.
				\end{itemize}
					
				As the matter of fact, if we note that these two operators $ a^{\dagger } $ and $ a $ need satisfy the relationship
				\begin{equation}
					a a^{\dagger } + a^{\dagger } a = \mathds{1} _{\mathrm{TC}} \ ,
				\end{equation}
				in the TC and, so, they are such that
				\begin{equation}
					a^{\dagger } \left\vert 0 \right\rangle = \left\vert 1 \right\rangle \ , \quad a \left\vert 1 \right\rangle = \left\vert 0 \right\rangle \quad \textnormal{and} \quad a^{\dagger } \left\vert 1 \right\rangle = a \left\vert 0 \right\rangle = 0 \cdot \left\vert 0 \right\rangle + 0 \cdot \left\vert 1 \right\rangle \ ,
				\end{equation}
				it becomes clear that all the operators defining the Hamiltonian (\ref{toric-hamil}) can also be expressed in terms of $ a^{\dagger } $ and $ a $: since the expression of these two operators (in terms of the same basis chosen on page \pageref{base-escolhida-page}) is
				\begin{equation*}
					a^{\dagger } =
					\begin{pmatrix}
						\ 0 & 0 \ \\
						\ 1 & 0 \
					\end{pmatrix} \quad \textnormal{and} \quad a =
					\begin{pmatrix}
						\ 0 & 1 \ \\
						\ 0 & 0 \
					\end{pmatrix}
				\end{equation*}
				and this implies, for instance, that
				\begin{equation*}
					\mathsf{n} = a^{\dagger } a =
					\begin{pmatrix}
						\ 0 & 0 \ \\
						\ 0 & 1 \
					\end{pmatrix} \quad \textnormal{and} \quad \mathsf{h} = a a^{\dagger } =
					\begin{pmatrix}
						\ 1 & 0 \ \\
						\ 0 & 0 \
					\end{pmatrix} \ ,
				\end{equation*}
				we see that
				\begin{equation}
					\sigma ^{x} = a^{\dagger } + a \quad \textnormal{and} \quad \sigma ^{z} = a a ^{\dagger } - a^{\dagger } a \ ,
				\end{equation}
				further reinforcing the counting capacity of vertex and face operators, because QFT considers itself $ \mathsf{n} $ as a \emph{count} operator (i.e., $ \mathsf{n} $ counts how many particles are present in a given physical state).
				
			\subsubsection{Fusion rules}	
			
				When we analyse the identity operator $ \mathds{1} _{j} $ by the same perspective
				\begin{equation}
					\mathds{1} _{j} = \mathds{1} _{j} \circ \mathds{1} _{j} \label{producao-vacuo}
				\end{equation}
				that (\ref{id-prop}), it is easy to conclude that all this scheme of quasiparticle pair creation also extends to it. The only difference here is that the quasiparticle pair associated with it must be composed of two vacuum quasiparticles which, whenever necessary, will be denoted as type $ 1 $. This is what happens when, for instance, we need to list the so-called \emph{fusion rules}	between the quasiparticles that belong to $ \left\{ 1 , e , m , \epsilon \right\} $ in the vicinity of a single edge. By according to what we have seen so far, the fusion rules related to these elementary quasiparticles are given by
				\begin{equation}
					\begin{array}{clrrrrrrrrrrrrrrrrr}
						1 & \times & 1 & = & e & \times & e & = & m & \times & m & = & \epsilon & \times & \epsilon & = & 1 & , \\
						1 & \times & e & = & e & \times & 1 & = & m & \times & \epsilon & = & \epsilon & \times & m & = & e & , \\
						1 & \times & m & = & m & \times & 1 & = & e & \times & \epsilon & = & \epsilon & \times&  e & = & m & , & \textnormal{and} \\
						1 & \times & \epsilon & = & \epsilon & \times & 1 & = & e & \times & m & = & m & \times & e & = & \epsilon & ,
					\end{array} \label{TC-fusion-rules}
				\end{equation}
				characterizing the \emph{Abelian} behaviour that was imposed on the group (\ref{TC-abelian}).
				
			\subsubsection{Statistical description}
					
				In view of these quasiparticles are transportable without increasing the system energy, before we finally close this Section it is interesting to evaluate the statistics of these quasiparticles. And the most effective way to do this is by analysing what happens when one of these quasiparticles \emph{spin around} another.
				
				By the way, one of the first things we can use for this evaluation is the fact that
				\begin{equation}
					\bigl[ O^{x} _{\boldsymbol{\gamma } ^{\ast } _{1}} , O^{x} _{\boldsymbol{\gamma } ^{\ast } _{2}} \bigr] = \bigl[ O^{z} _{\boldsymbol{\gamma } _{1}} , O^{z} _{\boldsymbol{\gamma } _{2}} \bigr] = 0 \ , \label{comut-caminhos-1}
				\end{equation}
				where $ \boldsymbol{\gamma } ^{\ast } _{1} $, $ \boldsymbol{\gamma } ^{\ast } _{2} $, $ \boldsymbol{\gamma } _{1} $ and $ \boldsymbol{\gamma } _{2} $ are arbitrary paths. After all, if we consider an initial state where a pair of quasiparticles $ e $ arises by the action of $ O^{x} _{\boldsymbol{\gamma } ^{\ast } _{1}} $ on any vacuum state, and consider that one of them revolves around the other via a contractile $ \bar{\boldsymbol{\gamma }} ^{\ast } _{2} $, we can conclude that these quasiparticles behave like \emph{bosons} among themselves because (\ref{comut-caminhos-1}) implies that the state obtained at the end of this process is exactly the same as the initial one.
				
				Of course, this bosonic conclusion also extends to the situation where only two quasiparticles $ m $ appear. Yet, as when one path $ \boldsymbol{\gamma } _{3} $ intercepts another $ \boldsymbol{\gamma } ^{\ast } _{3} $ in a single point we have
				\begin{equation}
					O^{x} _{\boldsymbol{\gamma } ^{\ast } _{3}} \circ O^{z} _{\boldsymbol{\gamma } ^{ } _{3}} = - O^{z} _{\boldsymbol{\gamma } ^{ } _{3}} \circ O^{x} _{\boldsymbol{\gamma } ^{\ast } _{3}} \label{comut-caminhos-2} \ ,
				\end{equation}
				it is immediate to conclude that a non-bosonic interpretation arises when a quasiparticle $ e $ revolves around another one of type $ m $ via the shortest path and vice versa. In order to understand this non-bosonic behaviour, just take
				\begin{equation}
					\left\vert \xi _{ \mathsf{initial}} \right\rangle = O^{x} _{\boldsymbol{\gamma } ^{\ast } _{3}} \circ O^{z} _{\boldsymbol{\gamma } ^{ } _{3}} \left\vert \xi _{0} \right\rangle \label{initial-anyon}
				\end{equation}
				as an initial physical state where we have only two quasiparticle pairs due to the action of $ O^{x} _{\boldsymbol{\gamma } ^{\ast } _{3}} $ and $ O^{z} _{\boldsymbol{\gamma } _{3}} $ in any vacuum states. After all, when a single quasiparticle $ e $ rotates around a \emph{single} of type $ m $ via a contractile $ \bar{\boldsymbol{\gamma }} _{4} $, we have
				\begin{eqnarray*}
					\left\vert \xi _{ \mathsf{final}} \right\rangle \negthickspace & = & \negthickspace O^{z} _{\bar{\boldsymbol{\gamma }} ^{ } _{4}} \left\vert \xi _{ \mathsf{initial}} \right\rangle \\
					& = & \negthickspace O^{z} _{\bar{\boldsymbol{\gamma }} ^{ } _{4}} \circ \bigl( O^{x} _{\boldsymbol{\gamma } ^{\ast } _{3}} \circ O^{z} _{\boldsymbol{\gamma } ^{ } _{3}} \bigr) \left\vert \xi _{0} \right\rangle = \bigl( O^{z} _{\bar{\boldsymbol{\gamma }} ^{ } _{4}} \circ O^{x} _{\boldsymbol{\gamma } ^{\ast } _{3}} \bigr) \circ O^{z} _{\boldsymbol{\gamma } ^{ } _{3}} \left\vert \xi _{0} \right\rangle \\
					& = & \negthickspace - O^{x} _{\boldsymbol{\gamma } ^{\ast } _{3}} \circ \bigl( O^{z} _{\bar{\boldsymbol{\gamma }} ^{ } _{4}} \circ O^{z} _{\boldsymbol{\gamma } ^{ } _{3}} \bigr) \left\vert \xi _{0} \right\rangle = - O^{x} _{\boldsymbol{\gamma } ^{\ast } _{3}} \circ O^{z} _{\boldsymbol{\gamma } ^{ } _{3}} \circ \left( O^{z} _{\bar{\boldsymbol{\gamma }} ^{ } _{4}} \left\vert \xi _{0} \right\rangle \right) \ .
				\end{eqnarray*}
				Thus, as (\ref{contract-condition}) and the contractility of $ \bar{\boldsymbol{\gamma }} _{4} $ allows to recognize that
				\begin{equation*}
					\prod _{p \in \mathsf{f} _{4}} B_{f} \circ O^{z} _{\bar{\boldsymbol{\gamma }} ^{ } _{4}} = O^{z} _{\bar{\boldsymbol{\gamma }} ^{ } _{4}} \circ \prod _{p \in \mathsf{f} _{4}} B_{f} = \mathds{1} _{\mathrm{TC}} \Rightarrow O^{z} _{\bar{\boldsymbol{\gamma }} ^{ } _{4}} \equiv \prod _{p \in \mathsf{f} _{4}} B_{f}
				\end{equation*}
				(where $ \mathsf{f} _{4} $ is the \emph{maximal} set containing the faces that are enclosed by $ \bar{\boldsymbol{\gamma }} _{4} $), it is clear that, thanks to (\ref{protege}), we have
				\begin{equation}
					\left\vert \xi _{ \mathsf{final}} \right\rangle = - O^{x} _{\boldsymbol{\gamma } ^{\ast } _{3}} \circ O^{z} _{\boldsymbol{\gamma } ^{ } _{3}} \circ \left( O^{z} _{\boldsymbol{\gamma } ^{ } _{4}} \left\vert \xi _{0} \right\rangle \right) = - O^{x} _{\boldsymbol{\gamma } ^{\ast } _{3}} \circ O^{z} _{\boldsymbol{\gamma } ^{ } _{3}} \left\vert \xi _{0} \right\rangle = - \left\vert \xi _{\mathsf{initial}} \right\rangle \ ; \label{comportamento-anyonico}
				\end{equation}
				i.e., a quasiparticle $ e $, when revolving around another type $ m $ and vice versa, cannot be interpreted as bosonic.
				
				In reason of the minus sign in (\ref{comportamento-anyonico}), two important comments can be made, and the first one follows when we take the case where two quasiparticles form a single dyon: after all, because $ \bar{\boldsymbol{\gamma }} ^{ } _{4} $ can be seen as the boundary of a single face, the rotation of these two quasiparticles around each other can be interpreted as the rotation of this dyon around itself. Thus, this minus sign in (\ref{comportamento-anyonico}) implies that a dyon is a fermion \cite{pachos}.
					
				The second comment is due to the fact that the rotation of a quasiparticle around another is equivalent to a \emph{double exchange} operation; in plain English, this rotation is equivalent to an operation in which these quasiparticles change places with each other and then return to their original positions due to a new exchange. Therefore, as in the more general case (where these two quasiparticles do not define a dyon) this minus sign in (\ref{comportamento-anyonico}) associates the statistic of these quasiparticles with a phase $ e^{i \pi / 2} $ that does not identify either with $ 0 $ (bosons) or $ 1 $ (fermions), we can conclude that quasiparticles $ e $ and $ m $ behave like anyons (spin-$ 1/4 $) in relative to each other.
				
			\subsubsection{One additional comment}
			
				Once it is possible to transport and, therefore, cause one quasiparticle to rotate around another, one thing is certain: a quasiparticle $ e $ will never collide with another of type $ m $, and it is by cause this observation that the great argument arises in favour of a distinction between two types ($ e $ and $ m $) of quasiparticles.
				
				However, as the minimum distance that can exist between them occurs when both define a single dyon, it is at this point that we need to clarify the reasons that led to their nomenclatures: after all, as the concept of anyon arose from the advent of the \emph{Aharonov-Bohm Effect} \cite{aha-bohm} (because it is possible to realize such anyonic statistics for systems where one electric particle rotates around one punctual magnetic field on a two-dimensional surface \cite{lerda-anyons}), it was possible to baptize the quasiparticles which are detectable in the vertices as type $ e $ (\underline{e}lectric), while those that are detectable in the faces ended up being denoted as type $ m $ (\underline{m}agnetic).
		
	\section{Three-dimensional considerations}
		
		Although there are already some generalizations of the TC (as is the case of Quantum Double Models (QDM)), the presentation we have made so far allows us to construct others, quite simple, which clearly bring the TC as one of their particular cases. One of these generalizations occurs, for instance, when we use the same TC encoding in a model that uses a \emph{three-dimensional} lattice $ \mathcal{L} _{3} $: scilicet, by
		\begin{itemize}
			\item associating a single qubit with each one of the $ \mathcal{L} _{3} $ edges, and
			\item defining physical operators capable of measuring the properties of this new model.
		\end{itemize}
		
		Of course, it is very comfortable to think that $ \mathcal{L} _{3} $ can discretize a manifold $ \mathcal{M} _{3} $ that, for instance, is identifiable as a three-dimensional torus $ \mathcal{T} _{3} $; and, in fact, this will be exactly one of the things we will do throughout this Section, since, for convenience, the manifold discretization we will adopt here will be \emph{cubic}. However, before attempting to understand a model defined on the discretization of $ \mathcal{T} _{3} $ (or of any other three-dimensional manifold), it is essential to understand what characteristics are common to all these models.
			
		\subsection{Similarities and differences}
		
			In spite of there are several possibilities for defining this new model on $ \mathcal{L} _{3} $, one thing is certain here: if we use the same logic that is used for defining the TC, it becomes very natural to assume that the Hamiltonian operator of this new three-dimensional code ($ 3 $DC) is given by
			\begin{equation}
				H_{\mathrm{3DC}} = - \sum _{v} A_{v} - \sum _{f} B_{f} \ , \label{toric-3d-hamil}
			\end{equation}
			where
			\begin{equation}
				A_{v} = \prod _{j \in S_{v}} \sigma ^{x} _{j} \quad \textnormal{and} \quad B_{f} = \prod _{j \in S_{f}} \sigma ^{z} _{j} \ . \label{$ 3 $-DC-operators}
			\end{equation}
			This natural assumption can be interpreted as a correspondence principle that must be imposed between the TC and $ 3 $DC, \textquotedblleft analogous\textquotedblright \hspace*{0.01cm} to that imposed between Newtonian and Quantum Mechanics, for instance.
			
			As a matter of fact, despite the undeniable similarity between this $ H_{\mathrm{3DC}} $ and the Hamiltonian (\ref{toric-hamil}), it is important to note that there is a difference between them that, although quite subtle, has a great influence on the behaviour of excitations that are measurable in the $ 3 $DC. After all, whereas the operators that make up (\ref{toric-hamil}) \emph{always} act effectively on four $ \mathcal{L} _{2} $ edges, in the $ 3 $DC this does not necessarily happen. The reason for this is that, regardless of the boundary conditions of $ \mathcal{L} _{3} $, each face and each vertex that make up the interior of this lattice are structured by \emph{four} and \emph{six} edges, respectively.
				
			\subsubsection{The vertex excitations}
				
				Before we explore the consequences of this last observation, it is also important to note that, when we restrict ourselves to a $ 3$DC where
				\begin{equation}
					\dim S_{f} = 4 \quad \textnormal{and} \quad \dim S_{v} = 6 \ , \label{condition-numbers}
				\end{equation}
				some of the main TC characteristics remain preserved, among which we can highlight:
				\begin{itemize}
					\item[(i)] the commutativity among these vertex and face operators (\ref{$ 3 $-DC-operators}), as well as their counting properties;
					\item[(ii)] the expression of the ground state energy as
					\begin{equation*}
						E_{0} = - \left( N_{v} + N_{f} \right) \ ,
					\end{equation*}
					where $ N_{v} $ and $ N_{f} $ are the respective quantities of vertices and faces that define $ \mathcal{L} _{3} $; and
					\item[(iii)] the fact that one of its vacuum states has exactly the same expression
					\begin{equation}
						\bigl\vert \xi ^{\left( 1 \right) } _{0} \bigr\rangle = \frac{1}{\sqrt{2}} \prod _{v} \left( \mathds{1} _{v} + A_{v} \right) \underbrace{ \left\vert 0 \right\rangle \otimes \ldots \otimes \left\vert 0 \right\rangle } _{N_{e} \textnormal{times}} \label{vacuo-3d-1}
					\end{equation}
					of the TC, where $ N_ {e} $ refers to the number of edges that define $ \mathcal{L} _{3} $.
				\end{itemize}
				
				Another familiar characteristic of this $ 3 $DC is associated with the most elementary vertex excitations. After all, since a vacuum state
				\begin{equation}
					\left\vert \xi ^{\prime \prime } \right\rangle = \sigma ^{z} _{k} \bigl\vert \xi ^{ \left( 1 \right) } _{0} \bigr\rangle \label{nao-vacuo-3d-exemplo-1}
				\end{equation}
				does not satisfy \emph{two} of the constraints
				\begin{equation}
					A_{v} \bigl\vert \xi _{0} \bigr\rangle = \bigl\vert \xi _{0} \bigr\rangle \quad \textnormal{and} \quad B_{f} \bigl\vert \xi _{0} \bigr\rangle = \bigl\vert \xi _{0} \bigr\rangle \label{protege-tri}
				\end{equation}
				that need to be satisfied by a vacuum state $ \bigl\vert \xi _{0} \bigr\rangle $, it is become clear that the energy of (\ref{nao-vacuo-3d-exemplo-1}) will also be equal to
				\begin{equation*}
					E_{1} = E_{0} + 4 \ ,
				\end{equation*}
				such as already happens in (\ref{nao-vacuo-exemplo-2}). Therefore, when we note that
				\begin{itemize}
					\item $ \sigma ^{z} _{k} $ can still create detectable excitations on the two vertices that enclose a $ k $-th edge of $ \mathcal{L} _{3} $ and, thence,
					\item these excitations are transportable only by the same operators already defined in (\ref{caminho-z}),
				\end{itemize}
				we can conclude not only that these excitations are bosons: we can also conclude that these excitations are exactly the same quasiparticles $ e $ already related to the TC.
				
			\subsubsection{The face excitations}
					
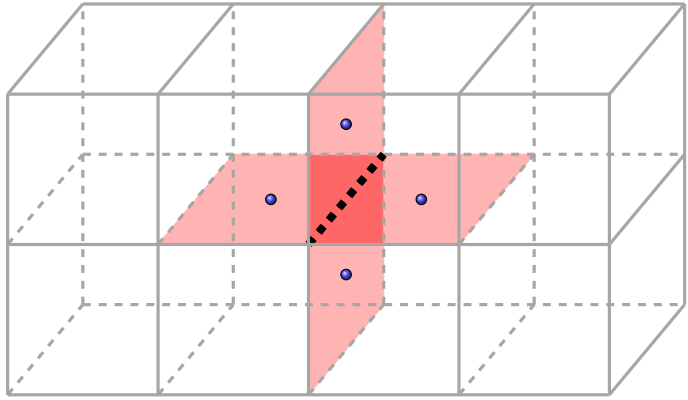
\begin{figure}[!t]
					\begin{center}
						\tikzstyle myBG=[line width=3pt,opacity=1.0]
						\newcommand{\drawLinewithBG}[2]
						{
							\draw[gray!70,very thick] (#1) -- (#2);
						}
						\newcommand{\drawDashedLinewithBG}[2]
						{
							\draw[gray!70,dashed,very thick] (#1) -- (#2);
						}
						\newcommand{\drawDashedLinewithBGV}[2]
						{
							\draw[gray!70,dashed,very thick] (#1) -- (#2);
						}
						\newcommand{\drawExcitedLinewithBGV}[2]
						{
							\draw[black,dashed,myBG]  (#1) -- (#2);
							\draw[dashed,very thick] (#1) -- (#2);
						}
						\newcommand{\graphLinesHorizontal}
						{
							% miolo -- linhas horizontais
							\drawDashedLinewithBG{1,1.2}{9,1.2};
							\drawDashedLinewithBG{1,3.2}{9,3.2};
							% miolo -- linhas verticais
							\drawDashedLinewithBG{1,1.2}{1,5.2};
							\drawDashedLinewithBG{3,1.2}{3,5.2};
							\drawDashedLinewithBG{5,1.2}{5,5.2};
							\drawDashedLinewithBG{7,1.2}{7,5.2};
							% miolo -- linhas transversas
							\drawDashedLinewithBG{0,0}{1,1.2};
							\drawDashedLinewithBG{2,0}{3,1.2};
							\drawDashedLinewithBG{4,0}{5,1.2};
							\drawDashedLinewithBG{6,0}{7,1.2};
							\drawDashedLinewithBG{0,2}{1,3.2};
							\drawDashedLinewithBGV{2,2}{3,3.2};
							\drawExcitedLinewithBGV{4,2}{5,3.2};
							\drawDashedLinewithBGV{6,2}{7,3.2};
							% envelope da rede -- linhas horizontais
							\drawLinewithBG{0,0}{8,0};
							\drawLinewithBG{0,2}{8,2};
							\drawLinewithBG{0,4}{8,4};
							\drawLinewithBG{1,5.2}{9,5.2};
							% envelope -- linhas verticais
							\drawLinewithBG{0,0}{0,4};
							\drawLinewithBG{2,0}{2,4};
							\drawLinewithBG{4,0}{4,4};
							\drawLinewithBG{6,0}{6,4};
							\drawLinewithBG{8,0}{8,4};
							\drawLinewithBG{9,1.2}{9,5.2};
							% envelope -- linhas transversas
							\drawLinewithBG{8,0}{9,1.2};
							\drawLinewithBG{8,2}{9,3.2};
							\drawLinewithBG{0,4}{1,5.2};
							\drawLinewithBG{2,4}{3,5.2};
							\drawLinewithBG{4,4}{5,5.2};
							\drawLinewithBG{6,4}{7,5.2};
							\drawLinewithBG{8,4}{9,5.2};
						}
						\begin{tikzpicture}
							\draw[color=red!30,fill=red!30] (4,0) -- (4,4) -- (5,5.2) -- (5,1.2) -- cycle;
							\draw[color=red!30,fill=red!30] (2,2) -- (3,3.2) -- (7,3.2) -- (6,2) -- cycle;
							\draw[color=red!30,fill=red!60] (4,2) -- (5,2) -- (5,3.2) -- (4,3.2) -- cycle;
							\graphLinesHorizontal;
							\draw[color=black,ball color=blue!70] (3.5,2.6) circle (0.07);
							\draw[color=black,ball color=blue!70] (5.5,2.6) circle (0.07);
							\draw[color=black,ball color=blue!70] (4.5,3.6) circle (0.07);
							\draw[color=black,ball color=blue!70] (4.5,1.6) circle (0.07);
						\end{tikzpicture}
					\end{center}
					\caption{Piece of the cubic lattice $ \mathcal{L} _{3} $ that supports the $ 3 $DC illustrating the presence of four face excitations (rose-coloured sectors around a single edge) due to the action of a single $ \sigma ^{x} _{k} $ on a vacuum state.}
					\label{asa-quasiplaca}
				\end{figure}

				However, when we look to another non-vacuum state
				\begin{equation}
					\left\vert \xi ^{\prime } \right\rangle = \sigma ^{x} _{k} \bigl\vert \xi ^{ \left( 1 \right) } _{0} \bigr\rangle \label{nao-vacuo-3d-exemplo-2}
				\end{equation}
				something a little different happens, because we are restricted to a region where (\ref{condition-numbers}) is valid and it implies that \emph{all} the $ \mathcal{L} _{3} $ edges belong to the intersection of \emph{four} faces. Thus, as the number of vacuum constraints that are violated by (\ref{nao-vacuo-3d-exemplo-2}) becomes equal to \emph{four}, we conclude that, unlike what happens in the TC, the action of a single $ \sigma ^{x} _{k} $ on the vacuum eigenstate (\ref{vacuo-3d-1}) creates \emph{four} simultaneous excitations around the $ k $-th edge, as shown in Figure \ref{asa-quasiplaca}.
				
				Moreover, due to this simultaneous creation of four excitations around a single edge, it is important to note that, in addition to eigenstate (\ref{nao-vacuo-3d-exemplo-2}) has an eigenenergy
				\begin{equation*}
					E_{2} = E_{0} + 8
				\end{equation*}
				that is clearly greater than $ E_{1} $, both transport and annihilation of these $ 3 $DC excitations need to be done in a slightly different way than the previous one. Of course, as $ \sigma ^{x} _{k} $ is defined in the same way as (\ref{sigma-composto}), it is good to note that these four excitations can be perfectly annihilated all together: this happens when $ \sigma ^{x} _{k} $ acts on $ \left\vert \xi ^{\prime } \right\rangle $, because
				\begin{equation*}
					\sigma ^{x} _{k} \left\vert \xi ^{\prime } \right\rangle = \sigma ^{x} _{k} \circ \left( \sigma ^{x} _{k} \bigl\vert \xi ^{ \left( 1 \right) } _{0} \bigr\rangle \right) = \left( \sigma ^{x} _{k} \circ \sigma ^{x} _{k} \right) \bigl\vert \xi ^{ \left( 1 \right) } _{0} \bigr\rangle = \bigl\vert \xi ^{ \left( 1 \right) } _{0} \bigr\rangle \ . 
				\end{equation*}
				
\begin{figure}[!t]
					\begin{center}
						\tikzstyle myBG=[line width=3pt,opacity=1.0]
						\newcommand{\drawLinewithBG}[2]
						{
							\draw[gray!70,very thick] (#1) -- (#2);
						}
						\newcommand{\drawDashedLinewithBG}[2]
						{
							\draw[gray!70,dashed,very thick] (#1) -- (#2);
						}
						\newcommand{\drawDashedLinewithBGV}[2]
						{
							\draw[gray!70,dashed,very thick] (#1) -- (#2);
						}
						\newcommand{\drawExcitedLinewithBGV}[2]
						{
							\draw[black,dashed,myBG]  (#1) -- (#2);
							\draw[dashed,very thick] (#1) -- (#2);
						}
						\newcommand{\graphLinesHorizontal}
						{
							% miolo -- linhas horizontais
							\drawDashedLinewithBG{1,1.2}{9,1.2};
							\drawDashedLinewithBG{1,3.2}{9,3.2};
							% miolo -- linhas verticais
							\drawDashedLinewithBG{1,1.2}{1,5.2};
							\drawDashedLinewithBG{3,1.2}{3,5.2};
							\drawDashedLinewithBG{5,1.2}{5,5.2};
							\drawDashedLinewithBG{7,1.2}{7,5.2};
							% miolo -- linhas transversas
							\drawDashedLinewithBG{0,0}{1,1.2};
							\drawDashedLinewithBG{2,0}{3,1.2};
							\drawDashedLinewithBG{4,0}{5,1.2};
							\drawDashedLinewithBG{6,0}{7,1.2};
							\drawDashedLinewithBG{0,2}{1,3.2};
							\drawDashedLinewithBGV{2,2}{3,3.2};
							\drawExcitedLinewithBGV{4,2}{5,3.2};
							\drawExcitedLinewithBGV{6,2}{7,3.2};
							% envelope da rede -- linhas horizontais
							\drawLinewithBG{0,0}{8,0};
							\drawLinewithBG{0,2}{8,2};
							\drawLinewithBG{0,4}{8,4};
							\drawLinewithBG{1,5.2}{9,5.2};
							% envelope -- linhas verticais
							\drawLinewithBG{0,0}{0,4};
							\drawLinewithBG{2,0}{2,4};
							\drawLinewithBG{4,0}{4,4};
							\drawLinewithBG{6,0}{6,4};
							\drawLinewithBG{8,0}{8,4};
							\drawLinewithBG{9,1.2}{9,5.2};
							% envelope -- linhas transversas
							\drawLinewithBG{8,0}{9,1.2};
							\drawLinewithBG{8,2}{9,3.2};
							\drawLinewithBG{0,4}{1,5.2};
							\drawLinewithBG{2,4}{3,5.2};
							\drawLinewithBG{4,4}{5,5.2};
							\drawLinewithBG{6,4}{7,5.2};
							\drawLinewithBG{8,4}{9,5.2};
						}
						\begin{tikzpicture}
							\draw[color=red!30,fill=red!30] (4,0) -- (4,4) -- (5,5.2) -- (5,1.2) -- cycle;
							\draw[color=red!30,fill=red!30] (6,0) -- (6,4) -- (7,5.2) -- (7,1.2) -- cycle;
							\draw[color=red!30,fill=red!30] (2,2) -- (3,3.2) -- (5,3.2) -- (4,2) -- cycle;
							\draw[color=red!30,fill=red!30] (6,2) -- (7,3.2) -- (9,3.2) -- (8,2) -- cycle;
							\draw[color=red!30,fill=red!60] (4,2) -- (5,3.2) -- (4,3.2) -- cycle;
							\draw[color=red!30,fill=red!60] (6,2) -- (7,2) -- (7,3.2) -- cycle;
							\graphLinesHorizontal;
							\draw[color=black,ball color=blue!70] (3.5,2.6) circle (0.07);
							\draw[color=black,ball color=blue!70] (7.5,2.6) circle (0.07);
							\draw[color=black,ball color=blue!70] (4.5,3.6) circle (0.07);
							\draw[color=black,ball color=blue!70] (4.5,1.6) circle (0.07);
							\draw[color=black,ball color=blue!70] (6.5,3.6) circle (0.07);
							\draw[color=black,ball color=blue!70] (6.5,1.6) circle (0.07);
						\end{tikzpicture}
					\end{center}
					\caption{Considering that the situation shown in Figure \ref{asa-quasiplaca} is associated with time $ t_{0} $, the actuation of a single $ \sigma ^{x} _{j} $ at one of the lattice edges at time $ t > t_{0} $ is not able to transport any face excitation in the same way as occurs in the TC. After all, in addition to just one face excitation being transported by this operator, two more arise increasing the energy of the system.}
					\label{indissociation}
				\end{figure}
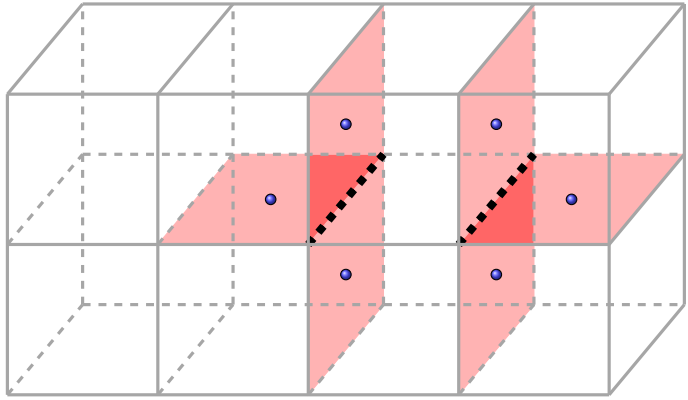

				Anyhow, if we wish to transport the excitations of (\ref{nao-vacuo-3d-exemplo-2}) by using an operator that does not identify with any \textquotedblleft teleport\textquotedblright 
				\begin{equation*}
					O^{x} _{jk} = \sigma ^{x} _{j} \circ \sigma ^{x} _{k} \ ,
				\end{equation*}
				this must be done in a way that is no longer as simple as previous ones. And in order to understand this, we should note that, when the $ j $-th edge of $ \mathcal{L} _{3} $ belongs to any face that contains the $ k $-th edge, only one of the four excitations created by a $ \sigma ^{x} _{j} $ will be able to annihilate a single one that was previously created by $ \sigma ^{x} _{k} $. As illustrated in Figure \ref{indissociation},
				this annihilation will occur precisely on the face (partially) bounded by these two edges. Thus, since, in general, the action of
				\begin{equation}
					O^{x} _{\Upsilon } = \prod _{j \in \Upsilon } \sigma ^{x} _{j} \label{energizator}
				\end{equation}
				on the eigenstate (\ref{nao-vacuo-3d-exemplo-2}) is not able to preserve the number of excitations associated with the set $ \Upsilon $ (which consist of edges that, two to two, belong to a single face), if we wish to transport these excitations through this lattice $ \mathcal{L} _{3} $ without increasing the system energy, this should be done in a smarter way.
				
			\subsubsection{A joint transport}
			
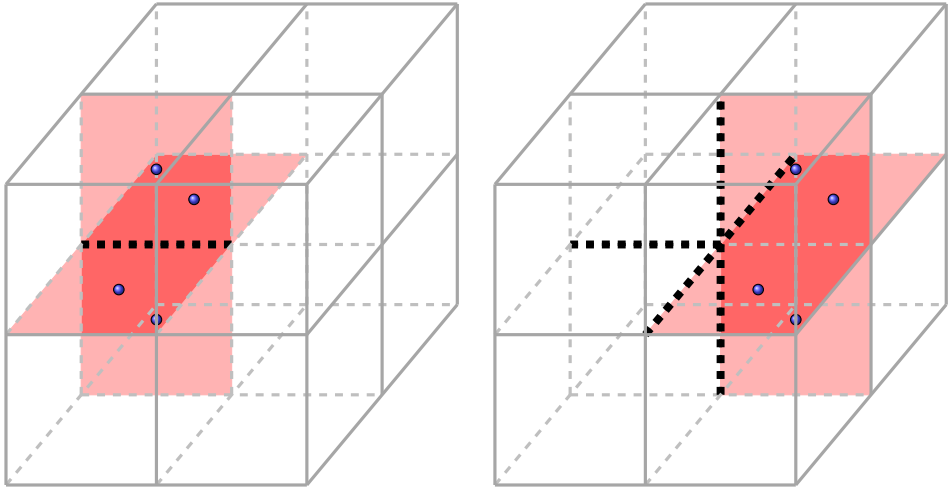
\begin{figure}[!t]
					\begin{center}
						\tikzstyle myBG=[line width=3pt,opacity=1.0]
						\newcommand{\drawLinewithBG}[2]
						{
							\draw[gray!70,very thick] (#1) -- (#2);
						}
						\newcommand{\drawDashedLinewithBG}[2]
						{
							\draw[gray!50,dashed,very thick] (#1) -- (#2);
						}
						\newcommand{\drawDashedLinewithBGV}[2]
						{
							\draw[gray!50,dashed,very thick] (#1) -- (#2);
						}
						\newcommand{\drawExcitedLinewithBGV}[2]
						{
							\draw[black,dashed,myBG]  (#1) -- (#2);
							\draw[dashed,very thick] (#1) -- (#2);
						}
						\newcommand{\graphLinesHorizontal}
						{
							% miolo -- linhas horizontais
							\drawDashedLinewithBG{1,1.2}{5,1.2};
							\drawDashedLinewithBG{7.5,1.2}{11.5,1.2};
							\drawDashedLinewithBG{1,3.2}{5,3.2};
							\drawDashedLinewithBG{7.5,3.2}{11.5,3.2};
							\drawExcitedLinewithBGV{1,3.2}{3,3.2}
							\drawExcitedLinewithBGV{7.5,3.2}{9.5,3.2}
							\drawDashedLinewithBG{2,2.4}{6,2.4};
							\drawDashedLinewithBG{8.5,2.4}{12.5,2.4};
							\drawDashedLinewithBG{2,4.4}{6,4.4};
							\drawDashedLinewithBG{8.5,4.4}{12.5,4.4};
							% miolo -- linhas verticais
							\drawDashedLinewithBG{1,1.2}{1,5.2};
							\drawDashedLinewithBG{3,1.2}{3,5.2};
							\drawDashedLinewithBG{7.5,1.2}{7.5,5.2};
							\drawExcitedLinewithBGV{9.5,1.2}{9.5,5.2};
							\drawDashedLinewithBG{2,2.4}{2,6.4};
							\drawDashedLinewithBG{4,2.4}{4,6.4};
							\drawDashedLinewithBG{8.5,2.4}{8.5,6.4};
							\drawDashedLinewithBG{10.5,2.4}{10.5,6.4};
							% miolo -- linhas transversas
							\drawDashedLinewithBG{0,0}{2,2.4};
							\drawDashedLinewithBG{2,0}{4,2.4};
							\drawDashedLinewithBG{6.5,0}{8.5,2.4};
							\drawDashedLinewithBG{8.5,0}{10.5,2.4};
							\drawDashedLinewithBG{0,2}{2,4.4};
							\drawDashedLinewithBGV{2,2}{4,4.4};
							\drawDashedLinewithBGV{6.5,2}{8.5,4.4};
							\drawExcitedLinewithBGV{8.5,2}{10.5,4.4};
							% envelope da rede -- linhas horizontais
							\drawLinewithBG{0,0}{4,0};
							\drawLinewithBG{6.5,0}{10.5,0};
							\drawLinewithBG{0,2}{4,2};
							\drawLinewithBG{6.5,2}{10.5,2};
							\drawLinewithBG{0,4}{4,4};
							\drawLinewithBG{6.5,4}{10.5,4};
							\drawLinewithBG{1,5.2}{5,5.2};
							\drawLinewithBG{7.5,5.2}{11.5,5.2};
							\drawLinewithBG{2,6.4}{6,6.4};
							\drawLinewithBG{8.5,6.4}{12.5,6.4};
							% envelope -- linhas verticais
							\drawLinewithBG{0,0}{0,4};
							\drawLinewithBG{2,0}{2,4};
							\drawLinewithBG{4,0}{4,4};
							\drawLinewithBG{5,1.2}{5,5.2};
							\drawLinewithBG{6,2.4}{6,6.4};
							\drawLinewithBG{6.5,0}{6.5,4};
							\drawLinewithBG{8.5,0}{8.5,4};
							\drawLinewithBG{10.5,0}{10.5,4};
							\drawLinewithBG{11.5,1.2}{11.5,5.2};
							\drawLinewithBG{12.5,2.4}{12.5,6.4};
							% envelope -- linhas transversas
							\drawLinewithBG{4,0}{6,2.4};
							\drawLinewithBG{4,2}{6,4.4};
							\drawLinewithBG{0,4}{2,6.4};
							\drawLinewithBG{2,4}{4,6.4};
							\drawLinewithBG{4,4}{6,6.4};
							\drawLinewithBG{10.5,0}{12.5,2.4};
							\drawLinewithBG{10.5,2}{12.5,4.4};
							\drawLinewithBG{6.5,4}{8.5,6.4};
							\drawLinewithBG{8.5,4}{10.5,6.4};
							\drawLinewithBG{10.5,4}{12.5,6.4};
						}
						\begin{tikzpicture}
							\draw[color=red!30,fill=red!30] (1,1.2) -- (1,5.2) -- (3,5.2) -- (3,1.2) -- cycle;
							\draw[color=red!30,fill=red!30] (0,2) -- (2,4.4) -- (4,4.4) -- (2,2) -- cycle;
							\draw[color=red!30,fill=red!60] (2,2) -- (3,3.2) -- (3,4.4)  -- (2,4.4) -- (1,3.2) -- (1,2) -- cycle;
							\draw[color=red!30,fill=red!30] (9.5,1.2) -- (9.5,5.2) -- (11.5,5.2) -- (11.5,1.2) -- cycle;
							\draw[color=red!30,fill=red!30] (8.5,2) -- (10.5,4.4) -- (12.5,4.4) -- (10.5,2) -- cycle;
							\draw[color=red!30,fill=red!60] (10.5,2) -- (11.5,3.2) -- (11.5,4.4)  -- (10.5,4.4) -- (9.5,3.2) -- (9.5,2) -- cycle;
							\draw[color=black,ball color=blue!70] (2,2.2) circle (0.07);
							\draw[color=black,ball color=blue!70] (10.5,2.2) circle (0.07);
							\graphLinesHorizontal;
							\draw[color=black,ball color=blue!70] (1.5,2.6) circle (0.07);
							\draw[color=black,ball color=blue!70] (2.5,3.8) circle (0.07);
							\draw[color=black,ball color=blue!70] (2,4.2) circle (0.07);
							\draw[color=black,ball color=blue!70] (10,2.6) circle (0.07);
							\draw[color=black,ball color=blue!70] (11,3.8) circle (0.07);
							\draw[color=black,ball color=blue!70] (10.5,4.2) circle (0.07);
						\end{tikzpicture}
					\end{center}
					\caption{Considering that the situation exposed in Figure \ref{asa-quasiplaca} is also associated with an instant $ t_{0} $ where we have four excitations on the faces that share a $ k $-th edge (on left), the action of a single $ O^{x} _{\mathcal{S} _{v}} $, at the same time $ t > t_{0} $, on four edges that are complementary to the first (according to a vertex $ v $) is able to carry all the excitations to edge $ j = S_{v} \setminus \left( k \cup \mathcal{S} _{v} \right) $ (on right).}
					\label{transporte-quasiplaca}
				\end{figure}

				Albeit (\ref{energizator}) is not ideal to perform this transport, an interesting observation can already be made from it, provide as, instead of a $ \mathcal{S} $, we take a set $ \mathcal{S} _{v} $ containing only \emph{four} of the edges which define a $ v $-th vertex. In order to understand this, it is enough to consider the case in which the $ k $-th edge excited in (\ref{nao-vacuo-3d-exemplo-2}) belongs to $ S_{v} \setminus \mathcal{S} _{v} $. After all, as
				\begin{equation}
					O^{x} _{\mathcal{S} _{v}} \left\vert \xi ^{\prime } \right\rangle = O^{x} _{\mathcal{S} _{v}} \circ \sigma ^{x} _{k} \bigl\vert \xi ^{ \left( 1 \right) } _{0} \bigr\rangle \label{$ 3 $-DC-state-transporter}
				\end{equation}
				will no longer violate the same four vacuum constraints as (\ref{nao-vacuo-3d-exemplo-2}) (but it will continue to violate the other \emph{four} new vacuum constraints which are now related to the four faces that share the edge $ j = S_{v} \setminus \left( k \cup \mathcal{S} _{v} \right) $), it is immediate to conclude from
				\begin{equation*}
					O^{x} _{\mathcal{S} _{v}} \circ \sigma ^{x} _{k} = \sigma ^{x} _{j} \circ A_{v}
				\end{equation*}
				that all four previous face excitations were carried by $ O^{x} _{\mathcal{S} _{v}} $ to the adjacencies of the $ j $-th edge, as shown in Figure \ref{transporte-quasiplaca}.
				
\begin{figure}[!t]
					\begin{center}
						\tikzstyle myBG=[line width=3pt,opacity=1.0]
						\newcommand{\drawLinewithBG}[2]
						{
							\draw[gray!70,very thick] (#1) -- (#2);
						}
						\newcommand{\drawDashedLinewithBG}[2]
						{
							\draw[gray!70,dashed,very thick] (#1) -- (#2);
						}
						\newcommand{\drawDashedLinewithBGV}[2]
						{
							\draw[gray!70,dashed,very thick] (#1) -- (#2);
						}
						\newcommand{\drawExcitedDashedLinewithBGV}[2]
						{
							\draw[black,dashed,myBG]  (#1) -- (#2);
							\draw[dashed,very thick] (#1) -- (#2);
						}
						\newcommand{\graphLinesHorizontal}
						{
							% miolo -- linhas horizontais
							\drawDashedLinewithBG{1,1.2}{5,1.2};
							\drawDashedLinewithBG{7,1.2}{11,1.2};
							\drawDashedLinewithBG{1,3.2}{5,3.2};
							\drawDashedLinewithBG{7,3.2}{11,3.2};
							% miolo -- linhas verticais
							\drawDashedLinewithBG{1,1.2}{1,5.2};
							\drawDashedLinewithBG{3,1.2}{3,5.2};
							\drawDashedLinewithBG{7,1.2}{7,5.2};
							\drawDashedLinewithBG{9,1.2}{9,5.2};
							% miolo -- linhas transversas
							\drawDashedLinewithBG{0,0}{1,1.2};
							\drawDashedLinewithBG{2,0}{3,1.2};
							\drawDashedLinewithBG{6,0}{7,1.2};
							\drawDashedLinewithBG{8,0}{9,1.2};
							\drawDashedLinewithBG{0,2}{1,3.2};
							\drawExcitedDashedLinewithBGV{2,2}{3,3.2};
							\drawDashedLinewithBGV{6,2}{7,3.2};
%							\drawExcitedLinewithBGV{8,2}{8.5,2.6};
							\drawExcitedDashedLinewithBGV{8,2}{9,3.2};
							% envelope da rede -- linhas horizontais
							\drawLinewithBG{0,0}{4,0};
							\drawLinewithBG{6,0}{10,0};
							\drawLinewithBG{0,2}{4,2};
							\drawLinewithBG{6,2}{10,2};
							\drawLinewithBG{0,4}{4,4};
							\drawLinewithBG{6,4}{10,4};
							\drawLinewithBG{1,5.2}{5,5.2};
							\drawLinewithBG{7,5.2}{11,5.2};
							% envelope -- linhas verticais
							\drawLinewithBG{0,0}{0,4};
							\drawLinewithBG{2,0}{2,4};
							\drawLinewithBG{4,0}{4,4};
							\drawLinewithBG{5,1.2}{5,5.2};
							\drawLinewithBG{6,0}{6,4};
							\drawLinewithBG{8,0}{8,4};
							\drawLinewithBG{10,0}{10,4};
							\drawLinewithBG{11,1.2}{11,5.2};
							% envelope -- linhas transversas
							\drawLinewithBG{4,0}{5,1.2};
							\drawLinewithBG{4,2}{5,3.2};
							\drawLinewithBG{0,4}{1,5.2};
							\drawLinewithBG{2,4}{3,5.2};
							\drawLinewithBG{4,4}{5,5.2};
							\drawLinewithBG{10,0}{11,1.2};
							\drawLinewithBG{10,2}{11,3.2};
							\drawLinewithBG{6,4}{7,5.2};
							\drawLinewithBG{8,4}{9,5.2};
							\drawLinewithBG{10,4}{11,5.2};
						}
						\begin{tikzpicture}
							\draw[color=red!30,fill=red!30] (2,0) -- (2,4) -- (3,5.2) -- (3,1.2) -- cycle;
							\draw[color=red!30,fill=red!30] (0,2) -- (1,3.2) -- (5,3.2) -- (4,2) -- cycle;
							\draw[color=red!30,fill=red!60] (2,2) -- (3,2) -- (3,3.2) -- (2,3.2) -- cycle;
							\draw[color=blue!70,fill=yellow!40] (7.5,1.6) -- (7.5,3.6) -- (9.5,3.6) -- (9.5,1.6) -- cycle;
							\graphLinesHorizontal;
							\draw[color=black,ball color=blue!70] (1.5,2.6) circle (0.07);
							\draw[color=black,ball color=blue!70] (3.5,2.6) circle (0.07);
							\draw[color=black,ball color=blue!70] (2.5,3.6) circle (0.07);
							\draw[color=black,ball color=blue!70] (2.5,1.6) circle (0.07);
							\draw[color=black,ball color=blue!70] (7.5,2.6) circle (0.07);
							\draw[color=black,ball color=blue!70] (9.5,2.6) circle (0.07);
							\draw[color=black,ball color=blue!70] (8.5,3.6) circle (0.07);
							\draw[color=black,ball color=blue!70] (8.5,1.6) circle (0.07);
%				   		    \draw[color=black,ball color=blue!70] (8.5,2.6) circle (0.1);
						\end{tikzpicture}
					\end{center}
					\caption{On left we have the usual situation of the previous illustrations, where, around a single point, four simultaneous excitations were created by the action of a single $ \sigma ^{x} _{j} $. On right we have a new idealization where a yellow plaque (which is crossed by the $ j $-th edge) represents this same set of excitations.}
					\label{quasiplaca}
				\end{figure}
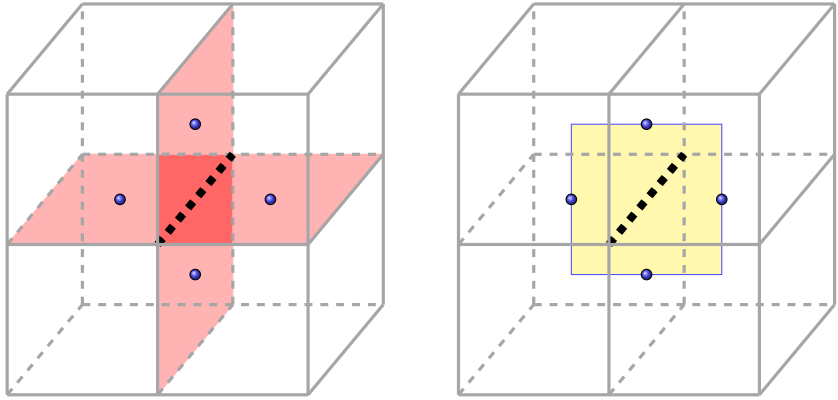

				In this fashion, since it is possible to transport these four face excitations by means of $ O^{x} _{\mathcal{S} _{v}} $ without increasing the system energy, it becomes possible to interpret this set of excitations as the three-dimensional generalization of a quasiparticle $ m $. Indeed, due to the fact that these four excitations are allocated on the four vertices that delimit one face on $ \mathcal{L} ^{\ast } _{3} $, it becomes natural to idealize this joint excitation as a kind of \textquotedblleft quasiplaque\textquotedblright , as illustrated in Figure \ref{quasiplaca}.
				As we shall see later, besides this \textquotedblleft quasiplaque\textquotedblright \hspace*{0.01cm} realization be quite useful for the understanding of the creation and transportation of these excitations, this realization also allows us to understand (in a much more comfortable way) how it is possible to conceive a dyon in the $ 3 $DC. 
					
		\subsection{Three-dimensional dyons}
				
			Before looking at the details of this three-dimensional dyon, it is essential to turn our attention to something that has great value to its conception: the operators
			\begin{equation}
				O^{z} _{\boldsymbol{\gamma }} = \prod _{j \in \boldsymbol{\gamma }} \sigma ^{z} _{j} \quad \textnormal{and} \quad O^{x} _{\boldsymbol{\gamma } ^{\prime }} = \prod _{v \in \boldsymbol{\gamma } ^{\prime }} O^{x} _{\mathcal{S} _{v}} \label{$ 3 $-DC-transporter}
			\end{equation}
			can transport quasiparticles $ e $ and \textquotedblleft quasiplaques\textquotedblright \hspace*{0.01cm} $ m $ respectively. Here, $ \boldsymbol{\gamma } $ and $ \boldsymbol{\gamma } ^{\prime } $ are the two paths that, also respectively, are composed of edges and vertices of $ \mathcal{L} _{3} $ which are two to two neighbours.
			
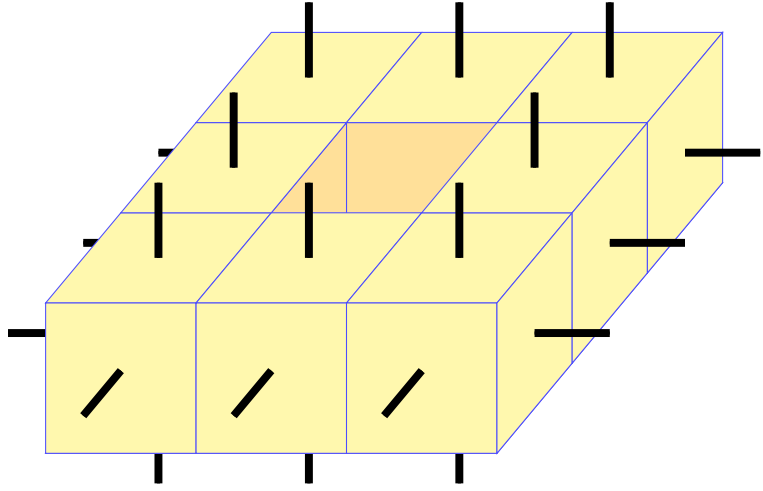
\begin{figure}[!t]
				\begin{center}
					\tikzstyle myBG=[line width=3pt,opacity=1.0]
					\newcommand{\drawLinewithBG}[2]
					{
						\draw[black,myBG]  (#1) -- (#2);
						\draw[black,very thick] (#1) -- (#2);
					}
					\newcommand{\drawDashedLinewithBG}[2]
					{
						\draw[gray!50,dashed,very thick] (#1) -- (#2);
					}
					\newcommand{\drawDashedLinewithBGV}[2]
					{
						\draw[gray!50,dashed,very thick] (#1) -- (#2);
					}
					\newcommand{\drawExcitedLinewithBGV}[2]
					{
						\draw[black,myBG]  (#1) -- (#2);
						\draw[black,very thick] (#1) -- (#2);
					}
					\newcommand{\drawExcitedDashedLinewithBGV}[2]
					{
						\draw[dashed,very thick] (#1) -- (#2);
					}
					\newcommand{\graphLinesFundo}
					{
						% envelope da rede -- linhas horizontais
						\drawLinewithBG{0,2}{4,2};
						\drawLinewithBG{1,3.2}{5,3.2};
						\drawLinewithBG{2,4.4}{6,4.4};
						% envelope -- linhas verticais
						\drawExcitedLinewithBGV{2,0}{2,1};
						\drawExcitedLinewithBGV{4,0}{4,1};
						\drawExcitedLinewithBGV{6,0}{6,1};
						% envelope -- linhas transversas
						\drawLinewithBG{4.6,3.1}{5.1,3.7};
					}
					\newcommand{\graphLinesFrente}
					{
						% envelope da rede -- linhas horizontais
						\drawLinewithBG{7,2}{8,2};
						\drawLinewithBG{8,3.2}{9,3.2};
						\drawLinewithBG{9,4.4}{10,4.4};
						% envelope -- linhas verticais
						\drawExcitedLinewithBGV{2,3}{2,4};
						\drawExcitedLinewithBGV{4,3}{4,4};
						\drawExcitedLinewithBGV{6,3}{6,4};
						\drawExcitedLinewithBGV{3,4.2}{3,5.2};
						\drawExcitedLinewithBGV{7,4.2}{7,5.2};
						\drawExcitedLinewithBGV{4,5.4}{4,6.4};
						\drawExcitedLinewithBGV{6,5.4}{6,6.4};
						\drawExcitedLinewithBGV{8,5.4}{8,6.4};
						% envelope -- linhas transversas
						\drawLinewithBG{1,0.9}{1.5,1.5};
						\drawLinewithBG{3,0.9}{3.5,1.5};
						\drawLinewithBG{5,0.9}{5.5,1.5};
					}
					\begin{tikzpicture}
						\graphLinesFundo;
						% placa horizontal -- inferior
						\draw[color=blue!70,fill=yellow!40] (0.5,0.4) -- (1.5,1.6) -- (3.5,1.6) -- (2.5,0.4) -- cycle;
						\draw[color=blue!70,fill=yellow!40] (2.5,0.4) -- (3.5,1.6) -- (5.5,1.6) -- (4.5,0.4) -- cycle;
						\draw[color=blue!70,fill=yellow!40] (4.5,0.4) -- (5.5,1.6) -- (7.5,1.6) -- (6.5,0.4) -- cycle;
						\draw[color=blue!70,fill=yellow!40] (1.5,1.6) -- (2.5,2.8) -- (4.5,2.8) -- (3.5,1.6) -- cycle;
						\draw[color=blue!70,fill=yellow!40] (5.5,1.6) -- (6.5,2.8) -- (8.5,2.8) -- (7.5,1.6) -- cycle;
						\draw[color=blue!70,fill=yellow!40] (2.5,2.8) -- (3.5,4.0) -- (5.5,4.0) -- (4.5,2.8) -- cycle;
						\draw[color=blue!70,fill=yellow!40] (4.5,2.8) -- (5.5,4.0) -- (7.5,4.0) -- (6.5,2.8) -- cycle;
						\draw[color=blue!70,fill=yellow!40] (6.5,2.8) -- (7.5,4.0) -- (9.5,4.0) -- (8.5,2.8) -- cycle;
						% miolo -- frente
						\draw[color=blue!70,fill=red!30!yellow!40] (4.5,2.8) -- (4.5,4.8) -- (6.5,4.8) -- (6.5,2.8) -- cycle;
						\draw[color=blue!70,fill=red!30!yellow!40] (3.5,1.6) -- (3.5,3.6) -- (4.5,4.8) -- (4.5,2.8) -- cycle;
						% placa horizontal -- superior
						\draw[color=blue!70,fill=yellow!40] (0.5,2.4) -- (1.5,3.6) -- (3.5,3.6) -- (2.5,2.4) -- cycle;
						\draw[color=blue!70,fill=yellow!40] (2.5,2.4) -- (3.5,3.6) -- (5.5,3.6) -- (4.5,2.4) -- cycle;
						\draw[color=blue!70,fill=yellow!40] (4.5,2.4) -- (5.5,3.6) -- (7.5,3.6) -- (6.5,2.4) -- cycle;
						\draw[color=blue!70,fill=yellow!40] (1.5,3.6) -- (2.5,4.8) -- (4.5,4.8) -- (3.5,3.6) -- cycle;
						\draw[color=blue!70,fill=yellow!40] (5.5,3.6) -- (6.5,4.8) -- (8.5,4.8) -- (7.5,3.6) -- cycle;
						\draw[color=blue!70,fill=yellow!40] (2.5,4.8) -- (3.5,6.0) -- (5.5,6.0) -- (4.5,4.8) -- cycle;
						\draw[color=blue!70,fill=yellow!40] (4.5,4.8) -- (5.5,6.0) -- (7.5,6.0) -- (6.5,4.8) -- cycle;
						\draw[color=blue!70,fill=yellow!40] (6.5,4.8) -- (7.5,6.0) -- (9.5,6.0) -- (8.5,4.8) -- cycle;
						% lateral -- frente
						\draw[color=blue!70,fill=yellow!40] (0.5,0.4) -- (0.5,2.4) -- (2.5,2.4) -- (2.5,0.4) -- cycle;
						\draw[color=blue!70,fill=yellow!40] (2.5,0.4) -- (2.5,2.4) -- (4.5,2.4) -- (4.5,0.4) -- cycle;
						\draw[color=blue!70,fill=yellow!40] (4.5,0.4) -- (4.5,2.4) -- (6.5,2.4) -- (6.5,0.4) -- cycle;
						\draw[color=blue!70,fill=yellow!40] (6.5,0.4) -- (6.5,2.4) -- (7.5,3.6) -- (7.5,1.6) -- cycle;
						\draw[color=blue!70,fill=yellow!40] (7.5,1.6) -- (7.5,3.6) -- (8.5,4.8) -- (8.5,2.8) -- cycle;
						\draw[color=blue!70,fill=yellow!40] (8.5,2.8) -- (8.5,4.8) -- (9.5,6.0) -- (9.5,4.0) -- cycle;
						\graphLinesFrente;
					\end{tikzpicture}
				\end{center}
				\caption{Here, we have a example of a closed tube that can be formed by \textquotedblleft quasiplaques\textquotedblright , provided that $ O^{x} _{\boldsymbol{\gamma }^{\prime }} $ only acts on four edges (in black) that define all the vertices that are points of a closed path $ \boldsymbol{\gamma } ^{\prime } $. Note that some of these edges do not appear in the figure: four of them cross the four central \textquotedblleft quasiplaques\textquotedblright \hspace*{0.01cm} (in gold), whereas the others are the ones that define $ \boldsymbol{\gamma } ^{\prime } $.}
				\label{toro-nao-retratil}
			\end{figure}

			In view of the observations made in the previous section, we do not need to give details of a vacuum eigenstate
			\begin{equation*}
				\bigl\vert \xi ^{\prime } _{0} \bigr\rangle = O^{z} _{\boldsymbol{\gamma }} \bigl\vert \xi ^{\left( 1 \right) } _{0} \bigr\rangle
			\end{equation*}
			that, for instance, arises when an operator $ O^{z} _{\boldsymbol{\gamma }} $ acts on a closed and contractile path $ \boldsymbol{\gamma } $. The only thing we need to do here is to evaluate an eigenstate such as
			\begin{equation}
				\bigl\vert \xi ^{\prime \prime } _{0} \bigr\rangle = O^{x} _{\boldsymbol{\gamma } ^{\prime }} \bigl\vert \xi ^{\left( 1 \right) } _{0} \bigr\rangle \ , \label{$ 3 $-DC-vacuo-x}
			\end{equation}
			which, owing to the action of $ O^{x} _{\boldsymbol{\gamma } ^{\prime }} $ on an arbitrary closed path $ \boldsymbol{\gamma } ^{\prime } $, comprises a vacuum structure that is identifiable as a \emph{closed tube} discretization, as shown in Figure \ref{toro-nao-retratil}.
			But, as behind this contractility, there is also the fact that
			\begin{equation*}
				O^{x} _{\boldsymbol{\gamma } ^{\prime }} = \prod _{v \in \boldsymbol{\gamma } ^{\prime }} O^{x} _{\mathcal{S} _{v}} = \prod _{v \in \boldsymbol{\gamma } ^{\prime }} \prod _{j \in S_{v}} \sigma ^{x} _{j} = \prod _{v \in \boldsymbol{\gamma } ^{\prime }} A_{v}
			\end{equation*}
			is always valid, we see that there is nothing very new here: the action of $ A_{v} $ continues to be interpreted as a mere gauge transformation which, by connecting (\ref{vacuo-3d-1}) to (\ref{$ 3 $-DC-vacuo-x}) and vice versa, makes it clear that these two eigenstates are modelling the same vacuum.
			
			Of course, the $ 3 $DC that we have been considering so far is too simplistic, since, for example, it does not contain vacuum eigenstates that are distinct from (\ref{vacuo-3d-1}), which arise from the action of the operators (\ref{$ 3 $-DC-transporter}) in three-dimensional lattices with more diversified boundary conditions. Yet, the expressions (\ref{$ 3 $-DC-transporter}) show us, for instance, how it is possible to conceive a dyon in the $ 3 $DC, because they show how it is possible to define compound operators
			\begin{equation*}
				O^{z} _{\boldsymbol{\gamma }} \circ O^{x} _{\boldsymbol{\gamma }} \quad \textnormal{and} \quad O^{x} _{\boldsymbol{\gamma }} \circ O^{z} _{\boldsymbol{\gamma }} \ .
			\end{equation*}
			After all, if we remember that one of the ways to create a dyon pair is via the action of a single
			\begin{equation}
				O^{x} _{\boldsymbol{\gamma } ^{\ast } _{1}} \circ O^{z} _{\boldsymbol{\gamma } _{1}} \label{TC-dyon-operator}
			\end{equation}
			on the vacuum eigenstate (\ref{vacuo-1}) where $ \boldsymbol{\gamma } ^{\ast } _{1} $ and $ \boldsymbol{\gamma } _{1} $ are two open paths whose ends coincide, it is immediate to conclude that this creation must be done by an operator
			\begin{equation}
				O^{x} _{\boldsymbol{\gamma }} \circ O^{z} _{\boldsymbol{\gamma }} \label{$ 3 $-DC-dyon-operator}
			\end{equation}
			acting on a $ 3 $DC vacuum eigenstate.
			
			However, in spite of this last operator creates two excitations which will denote by $ \epsilon ^{\prime } $, we still need to be careful before claiming that this pair of excitations actually matches a dyon pair in this three-dimensional model. After all, although (\ref{TC-dyon-operator}) really indicates that the simplest way to project a dyon pair on a TC (constrained to a $ \mathcal{L} _{2} $ embedded in $ \mathcal{L} _{3} $) is by means of a single operator
			\begin{equation}
				\sigma ^{x} _{j} \circ \sigma ^{z} _{j} \label{dyon-e-creation}
			\end{equation}
			acting on the same edge of (\ref{vacuo-1}), it is enough to see that, if we assume that a coupled excitation\footnote{Consisting of a pair composed by one quasiparticle $ e $ and one \textquotedblleft quasiplaque\textquotedblright \hspace*{0.01cm} $ m $ which, in addition to being arranged as short as possible, are transportable without increasing the system energy.} summarized in a single $ \epsilon ^{\prime } $ must be interpreted as a single dyon, it is immediate to realize that (\ref{dyon-e-creation}) does not create a dyon pair in the $ 3 $DC: what it does is just create one dyon plus one quasiparticle $ e $.
				
			Despite this seems to ruin any chance of the $ 3 $DC to correspond to the TC through a dimensional reduction procedure, an interesting result appears when we turn our attention to an operator
			\begin{equation}
				\sigma ^{z} _{j} \circ O^{\ x} _{\bar{\mathcal{S}} _{v}} \ , \label{$ 3 $-DC-dyon-creation-simplist}
			\end{equation}
			whose $ j $-th edge belongs to $ S_{v} $ but does not belong to the set $ \bar{\mathcal{S} _{v}} $ that is composed of four \emph{coplanar} edges. After all, as
			\begin{equation*}
				O^{x} _{\bar{\mathcal{S}} _{v}} \circ A_{v} = \sigma ^{x} _{j} \circ \sigma ^{x} _{k}
			\end{equation*}
			shows us that
			\begin{equation*}
				O^{x} _{\bar{\mathcal{S}} _{v}} \bigl\vert \xi ^{ \left( 1 \right) } _{0} \bigr\rangle = O^{x} _{\bar{\mathcal{S}} _{v}} \circ A_{v} \bigl\vert \xi ^{ \left( 1 \right) } _{0} \bigr\rangle = \sigma ^{x} _{j} \circ \sigma ^{x} _{k} \bigl\vert \xi ^{ \left( 1 \right) } _{0} \bigr\rangle
			\end{equation*}
			where $ j $ and $ k $ are indexing two opposite edges of a face, this result tells us that $ O^{x} _{\bar{\mathcal{S}} _{v}} $ is an operator that can create one \textquotedblleft quasiplaque\textquotedblright \hspace*{0.01cm} pair and, so, the operator (\ref{$ 3 $-DC-dyon-creation-simplist}) can create one pair of $ \epsilon ^{\prime } $.
				
			\subsubsection{A statistical comment}
				
			 	Albeit this conclusion indicates that a correspondence principle exists between the TC and $ 3 $DC if $ \epsilon ^{\prime } $ is actually a dyon, we must note two important things here. And the first one concerns to a statistical ambiguity that seems to be related to $ \epsilon ^{\prime } $ components: after all, rotate one quasiparticle $ e $ around a \textquotedblleft quasiplaque\textquotedblright \hspace*{0.01cm} $ m $ seems to imply that both are
				\begin{itemize}
					\item anyons with spin-$ 1/4 $, if the path chosen for this rotation cross the \textquotedblleft quasiplaque\textquotedblright \hspace*{0.01cm} $ m $, and
					\item bosons, otherwise.
				\end{itemize}											
				However, as this supposed dyon $ \epsilon ^{\prime } $ needs to keep it intact under any transportation, all this apparent ambiguity does not matter for its definition because the quasiparticle $ e $ that makes up $ \epsilon ^{\prime } $ always needs to cross a \textquotedblleft quasiplaque\textquotedblright \hspace*{0.01cm} $ m $ an odd number of times to complete any trajectory. In other words, as with the dyons associated to the TC, $ \epsilon ^{\prime } $ also behaves like a \emph{fermion} in relation to another $ \epsilon ^{\prime } $ and its fermionicity is a feature that is not shared by the other $ 3 $DC quasiparticles $ e $ and $ m $.
				
				In view this, the second observation comes in the form of a questioning: if quasiparticles $ e $ and $ m $ cannot cross each other in the TC, why is it reasonable to assume that this happens in the $ 3 $DC? And the best answer we can give to this questioning is that a dyon is a quasiparticle that \textquotedblleft happens by chance\textquotedblright \hspace*{0.01cm} in these models. That is, despite $ \epsilon $ has the status of a quasiparticle that is independent of the others in the two-dimensional TC (because it corresponds effectively to the quasiparticle that would be created by $ \sigma ^{y} _{j} $), nothing obliges the quasiparticles $ e $ and $ m $ to structure a dyon forever. Therefore, as the operator (\ref{$ 3 $-DC-dyon-operator}) (which is the three-dimensional generalization of (\ref{producao-epsilon})) creates a fermionic pair whose quasiparticles structuring these fermions can also be transported individually, it is valid to conclude that there is no statistical ambiguity related to $ \epsilon ^{\prime } $, and so that this quasiparticle is a $ 3 $DC dyon. 
				
			\subsubsection{An interesting comment}
			
				Although this does not seem to be the case in the TC, all these crossings are already common in this two-dimensional model. And in order to understand this, it is enough to analyse the case where, for example, a quasiparticle $ e $ needs to be transported through some edge whose adjacent faces already have two quasiparticles $ m $. After all, despite the two-dimensionality of $ \mathcal{L} _{2} $ prevents us from realizing the excitations created by $ \sigma ^{x} _{j} $ as the \textquotedblleft quasiplaque\textquotedblright \hspace*{0.01cm} that, for instance, is shown in Figure \ref{quasiplaca}, it is not absurd to realize the transport of $ e $ in terms of the crossing that this quasiparticle needs to make in the one-dimensional \textquotedblleft quasiplaque\textquotedblright \hspace*{0.01cm} previously created by $ \sigma ^{x} _{j} $.
				
				By virtue of this \textquotedblleft quasiplaque\textquotedblright \hspace*{0.01cm} realization, it becomes clear why the quasiparticles $ m $ are free in the TC but confined in the $ 3 $DC, because,
				\begin{itemize}
					\item whereas the situation illustrated in Figure \ref{perimeter-law}
					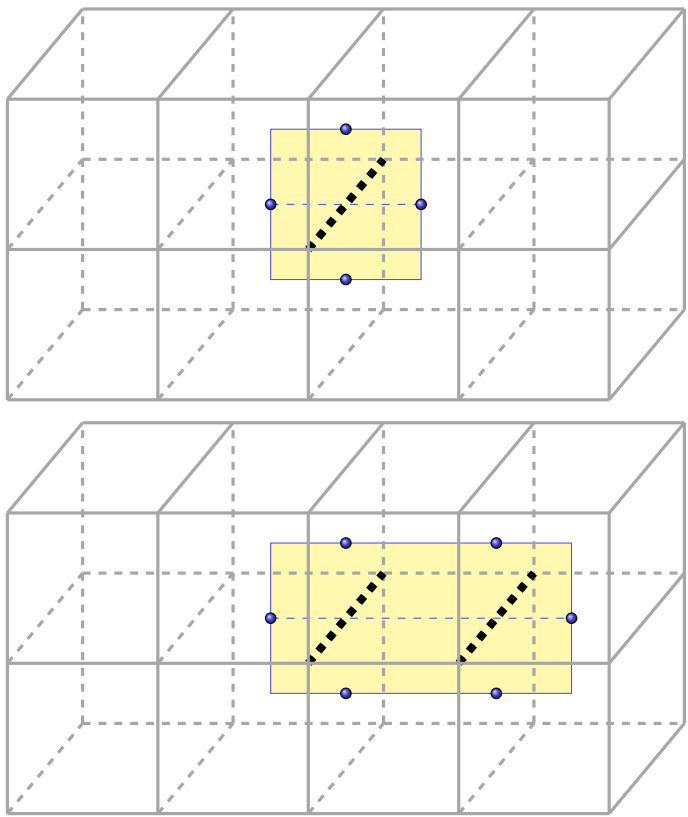
\begin{figure}[!t]
						\begin{center}
							\tikzstyle myBG=[line width=3pt,opacity=1.0]
							\newcommand{\drawLinewithBG}[2]
							{
								\draw[gray!70,very thick] (#1) -- (#2);
							}
							\newcommand{\drawDashedLinewithBG}[2]
							{
								\draw[gray!70,dashed,very thick] (#1) -- (#2);
							}
							\newcommand{\drawDashedLinewithBGV}[2]
							{
								\draw[gray!70,dashed,very thick] (#1) -- (#2);
							}
							\newcommand{\drawExcitedLinewithBGV}[2]
							{
								\draw[black,dashed,myBG]  (#1) -- (#2);
								\draw[dashed,very thick] (#1) -- (#2);
							}
							\newcommand{\graphLinesHorizontal}
							{
								% miolo -- linhas horizontais
								\drawDashedLinewithBG{1,1.2}{9,1.2};
								\drawDashedLinewithBG{1,3.2}{9,3.2};
								% miolo -- linhas verticais
								\drawDashedLinewithBG{1,1.2}{1,5.2};
								\drawDashedLinewithBG{3,1.2}{3,5.2};
								\drawDashedLinewithBG{5,1.2}{5,5.2};
								\drawDashedLinewithBG{7,1.2}{7,5.2};
								% miolo -- linhas transversas
								\drawDashedLinewithBG{0,0}{1,1.2};
								\drawDashedLinewithBG{2,0}{3,1.2};
								\drawDashedLinewithBG{4,0}{5,1.2};
								\drawDashedLinewithBG{6,0}{7,1.2};
								\drawDashedLinewithBG{0,2}{1,3.2};
								\drawDashedLinewithBGV{2,2}{3,3.2};
								\drawExcitedLinewithBGV{4,2}{5,3.2};
								\drawDashedLinewithBGV{6,2}{7,3.2};
								% envelope da rede -- linhas horizontais
								\drawLinewithBG{0,0}{8,0};
								\drawLinewithBG{0,2}{8,2};
								\drawLinewithBG{0,4}{8,4};
								\drawLinewithBG{1,5.2}{9,5.2};
								% envelope -- linhas verticais
								\drawLinewithBG{0,0}{0,4};
								\drawLinewithBG{2,0}{2,4};
								\drawLinewithBG{4,0}{4,4};
								\drawLinewithBG{6,0}{6,4};
								\drawLinewithBG{8,0}{8,4};
								\drawLinewithBG{9,1.2}{9,5.2};
								% envelope -- linhas transversas
								\drawLinewithBG{8,0}{9,1.2};
								\drawLinewithBG{8,2}{9,3.2};
								\drawLinewithBG{0,4}{1,5.2};
								\drawLinewithBG{2,4}{3,5.2};
								\drawLinewithBG{4,4}{5,5.2};
								\drawLinewithBG{6,4}{7,5.2};
								\drawLinewithBG{8,4}{9,5.2};
							}
							\begin{tikzpicture}
								\draw[color=blue!70,fill=yellow!40] (3.5,1.6) -- (3.5,3.6) -- (5.5,3.6) -- (5.5,1.6) -- cycle;
								\draw[color=blue!70,dashed] (3.5,2.6) -- (5.5,2.6);
								\graphLinesHorizontal;
								\draw[color=black,ball color=blue!70] (3.5,2.6) circle (0.07);
								\draw[color=black,ball color=blue!70] (5.5,2.6) circle (0.07);
								\draw[color=black,ball color=blue!70] (4.5,3.6) circle (0.07);
								\draw[color=black,ball color=blue!70] (4.5,1.6) circle (0.07);
							\end{tikzpicture}
						\end{center}
						\begin{center}
							\tikzstyle myBG=[line width=3pt,opacity=1.0]
							\newcommand{\drawLinewithBG}[2]
							{
								\draw[gray!70,very thick] (#1) -- (#2);
							}
							\newcommand{\drawDashedLinewithBG}[2]
							{
								\draw[gray!70,dashed,very thick] (#1) -- (#2);
							}
							\newcommand{\drawDashedLinewithBGV}[2]
							{
								\draw[gray!70,dashed,very thick] (#1) -- (#2);
							}
							\newcommand{\drawExcitedLinewithBGV}[2]
							{
								\draw[black,dashed,myBG]  (#1) -- (#2);
								\draw[dashed,very thick] (#1) -- (#2);
							}
							\newcommand{\graphLinesHorizontal}
							{
								% miolo -- linhas horizontais
								\drawDashedLinewithBG{1,1.2}{9,1.2};
								\drawDashedLinewithBG{1,3.2}{9,3.2};
								% miolo -- linhas verticais
								\drawDashedLinewithBG{1,1.2}{1,5.2};
								\drawDashedLinewithBG{3,1.2}{3,5.2};
								\drawDashedLinewithBG{5,1.2}{5,5.2};
								\drawDashedLinewithBG{7,1.2}{7,5.2};
								% miolo -- linhas transversas
								\drawDashedLinewithBG{0,0}{1,1.2};
								\drawDashedLinewithBG{2,0}{3,1.2};
								\drawDashedLinewithBG{4,0}{5,1.2};
								\drawDashedLinewithBG{6,0}{7,1.2};
								\drawDashedLinewithBG{0,2}{1,3.2};
								\drawDashedLinewithBGV{2,2}{3,3.2};
								\drawExcitedLinewithBGV{4,2}{5,3.2};
								\drawExcitedLinewithBGV{6,2}{7,3.2};
								% envelope da rede -- linhas horizontais
								\drawLinewithBG{0,0}{8,0};
								\drawLinewithBG{0,2}{8,2};
								\drawLinewithBG{0,4}{8,4};
								\drawLinewithBG{1,5.2}{9,5.2};
								% envelope -- linhas verticais
								\drawLinewithBG{0,0}{0,4};
								\drawLinewithBG{2,0}{2,4};
								\drawLinewithBG{4,0}{4,4};
								\drawLinewithBG{6,0}{6,4};
								\drawLinewithBG{8,0}{8,4};
								\drawLinewithBG{9,1.2}{9,5.2};
								% envelope -- linhas transversas
								\drawLinewithBG{8,0}{9,1.2};
								\drawLinewithBG{8,2}{9,3.2};
								\drawLinewithBG{0,4}{1,5.2};
								\drawLinewithBG{2,4}{3,5.2};
								\drawLinewithBG{4,4}{5,5.2};
								\drawLinewithBG{6,4}{7,5.2};
								\drawLinewithBG{8,4}{9,5.2};
							}
							\begin{tikzpicture}
								\draw[color=blue!70,fill=yellow!40] (3.5,1.6) -- (3.5,3.6) -- (7.5,3.6) -- (7.5,1.6) -- cycle;
								\draw[color=blue!70,dashed] (3.5,2.6) -- (7.5,2.6);
								\graphLinesHorizontal;
								\draw[color=black,ball color=blue!70] (3.5,2.6) circle (0.07);
								\draw[color=black,ball color=blue!70] (7.5,2.6) circle (0.07);
								\draw[color=black,ball color=blue!70] (4.5,3.6) circle (0.07);
								\draw[color=black,ball color=blue!70] (6.5,3.6) circle (0.07);
								\draw[color=black,ball color=blue!70] (4.5,1.6) circle (0.07);
								\draw[color=black,ball color=blue!70] (6.5,1.6) circle (0.07);
							\end{tikzpicture}
						\end{center}
						\caption{\textquotedblleft Quasiplaque\textquotedblright \hspace*{0.01cm} juxtaposition scheme that allows us to illustrate the perimeter law. After all, while in the figure above we have a single \textquotedblleft quasiplaque\textquotedblright \hspace*{0.01cm} with an energy equal to $ 4 $, in the figure below we have two juxtaposed \textquotedblleft quasiplaques\textquotedblright \hspace*{0.01cm} with an energy equal to $ 6 $: i.e., the total energy in both cases (as in all other cases) is exactly equal to the perimeter (number of edges) of the closed path that delimits this juxtaposition.}
						\label{perimeter-law}
					\end{figure}
					shows that the transport of the quasiparticles $ m $ along a dual path $ \boldsymbol{\gamma } ^{\ast } $ in $ \mathcal{L} _{3} $ (highlighted in dashed blue) display a \emph{perimeter law} for their energies (i.e., the energy of this transport is identified by the amount of dual edges enclosing the union of \textquotedblleft quasiplaques\textquotedblright \hspace*{0.01cm} that arise along $ \boldsymbol{\gamma } ^{\ast } $),
					\item this same law does \emph{not} apply to a model that is restricted to sublattice $ \mathcal{L} _{2} \subset \mathcal{L} _{3} $ that discretizes, for example, a submanifold $ \mathcal{M} _{2} \subset \mathcal{M} _{3} $.
				\end{itemize}
				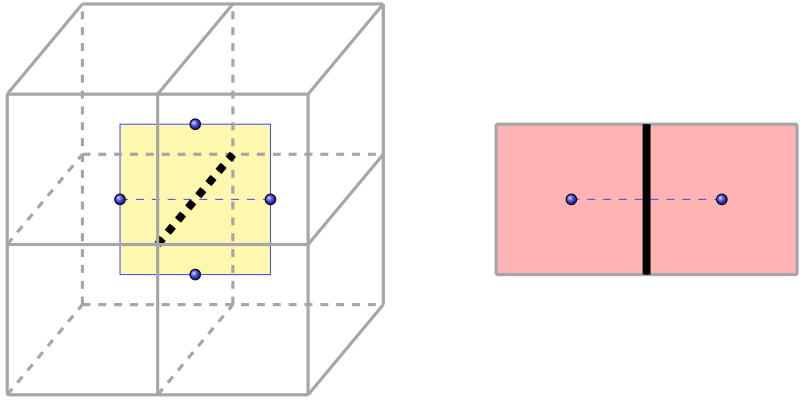
\begin{figure}[!t]
					\begin{center}
						\tikzstyle myBG=[line width=3pt,opacity=1.0]
						\newcommand{\drawLinewithBG}[2]
						{
							\draw[gray!70,very thick] (#1) -- (#2);
						}
						\newcommand{\drawDashedLinewithBG}[2]
						{
							\draw[gray!70,dashed,very thick] (#1) -- (#2);
						}
						\newcommand{\drawDashedLinewithBGV}[2]
						{
							\draw[gray!70,dashed,very thick] (#1) -- (#2);
						}
						\newcommand{\drawExcitedLinewithBGV}[2]
						{
							\draw[black,myBG]  (#1) -- (#2);
							\draw[very thick] (#1) -- (#2);
						}
						\newcommand{\drawExcitedDashedLinewithBGV}[2]
						{
							\draw[black,dashed,myBG]  (#1) -- (#2);
							\draw[dashed,very thick] (#1) -- (#2);
						}
						\newcommand{\graphLinesHorizontal}
						{
							% miolo -- linhas horizontais
							\drawDashedLinewithBG{1,1.2}{5,1.2};
							\drawDashedLinewithBG{1,3.2}{5,3.2};
							% miolo -- linhas verticais
							\drawDashedLinewithBG{1,1.2}{1,5.2};
							\drawDashedLinewithBG{3,1.2}{3,5.2};
							% miolo -- linhas transversas
							\drawDashedLinewithBG{0,0}{1,1.2};
							\drawDashedLinewithBG{2,0}{3,1.2};
							\drawDashedLinewithBG{0,2}{1,3.2};
							\drawExcitedDashedLinewithBGV{2,2}{3,3.2};
							% envelope da rede -- linhas horizontais
							\drawLinewithBG{0,0}{4,0};
							\drawLinewithBG{0,2}{4,2};
							\drawLinewithBG{0,4}{4,4};
							\drawLinewithBG{1,5.2}{5,5.2};
							\drawLinewithBG{6.5,1.6}{10.5,1.6};
							\drawLinewithBG{6.5,3.6}{10.5,3.6};
							% envelope -- linhas verticais
							\drawLinewithBG{0,0}{0,4};
							\drawLinewithBG{2,0}{2,4};
							\drawLinewithBG{4,0}{4,4};
							\drawLinewithBG{5,1.2}{5,5.2};
							\drawLinewithBG{6.5,1.6}{6.5,3.6};
							\drawExcitedLinewithBGV{8.5,1.6}{8.5,3.6};
							\drawLinewithBG{10.5,1.6}{10.5,3.6};
							% envelope -- linhas transversas
							\drawLinewithBG{4,0}{5,1.2};
							\drawLinewithBG{4,2}{5,3.2};
							\drawLinewithBG{0,4}{1,5.2};
							\drawLinewithBG{2,4}{3,5.2};
							\drawLinewithBG{4,4}{5,5.2};
						}
						\begin{tikzpicture}
							\draw[color=blue!70,fill=yellow!40] (1.5,1.6) -- (1.5,3.6) -- (3.5,3.6) -- (3.5,1.6) -- cycle;
							\draw[color=blue!70,dashed] (1.5,2.6) -- (3.5,2.6);
							\draw[color=red!30,fill=red!30] (6.5,1.6) rectangle (10.5,3.6);
							\draw[color=blue!70,dashed] (7.5,2.6) -- (9.5,2.6);
							\graphLinesHorizontal;
							\draw[color=black,ball color=blue!70] (1.5,2.6) circle (0.07);
							\draw[color=black,ball color=blue!70] (3.5,2.6) circle (0.07);
							\draw[color=black,ball color=blue!70] (2.5,3.6) circle (0.07);
							\draw[color=black,ball color=blue!70] (2.5,1.6) circle (0.07);
							\draw[color=black,ball color=blue!70] (7.5,2.6) circle (0.07);
							\draw[color=black,ball color=blue!70] (9.5,2.6) circle (0.07);
						\end{tikzpicture}
					\end{center}
					\caption{On left we use the same illustration used in Figure \ref{quasiplaca} to present a \textquotedblleft quasiplaque\textquotedblright \hspace*{0.01cm} but highlighting the intersection between it, which was created due to action of $ \sigma ^{x} _{j} $ on the $ j $-th edge, and the cutout of the horizontal plane that supports this edge. On right we see only the same cutout of this horizontal plane where two excitations that define a \textquotedblleft quasiplaque\textquotedblright \hspace*{0.01cm} are present, the only ones that exist in this two-dimensional environment. Note that all the excitations created by a single $ \sigma ^{x} _{j} $ in the two-dimensional model (which is obtained by a cut or a contraction of all planes parallel to a single plane discretizable by $ \mathcal{L} _{2} \subset \mathcal{L} _{3} $) are fully equivalent to the TC quasiparticles $ m $, which are transportable at no cost to the energy.}
					\label{deconfined-projection}
				\end{figure}
				In other words, this non-validity of the perimeter law can be understood in a very simple way, provided we observe that $ \mathcal{L} _{2} \subset \mathcal{L} _{3} $. After all, if we consider the situation of the transport of quasiparticles $ m $ from the perspective of $ \mathcal{L} _{3} $, this same transport, when restricted to the two-dimensional lattice $ \mathcal{L} _{2} $, can be realized in terms of cut (or projection) that is shown in Figure \ref{deconfined-projection}: that is, the transport situation of the quasiparticles $ m $, which, for instance, are associated with a model such as TC, can be achieved by restricting vertex and face operators (\ref{$ 3 $-DC-operators}) to act only on $ \mathcal{L} _{2} \subset \mathcal{L} _{3} $.
					
	\section{The three-dimensional Toric Code}
	
		Due to these positive aspects, and especially remembering that the definition of a dyon in the TC was made so that it was possible to define a model where \emph{all} its quasiparticles could be transported without any addition of energy, it is more than convenient to declare that we will do exactly the same with respect to $ 3 $DC: in plain English, we will declare that $ \epsilon ^{\prime } $ actually corresponds to a dyon $ \epsilon $, since the quasiparticle $ e $ and the \textquotedblleft quasiplaque\textquotedblright \hspace*{0.01cm} $ m $ that define it can also be transported without increasing the system energy.
		
		However, one thing that is very important to note here is that, in addition to the fusion rules related to all these quasiparticles $ e $ and \textquotedblleft quasiplaques\textquotedblright \hspace*{0.01cm} $ m $ are exactly the same as the TC (because they are created by the same operators), all results so far we have referred to the particular case of a $ 3 $DC that satisfies (\ref{condition-numbers}). And a good way of thinking about this is, for example, interpreting the $ 3 $DC as a model defined in a cubic lattice that (i) can be infinite or (ii) not infinite with periodic boundary conditions in all three directions. If this is indeed the case, the only thing we need to keep in mind is that
		\begin{itemize}
			\item whereas an $ 3 $DC constructed in the first lattice (which is infinite) has a unique vacuum state given by (\ref{vacuo-3d-1}),
			\begin{figure}[!t]
			\centering
			\includegraphics[viewport=330 30 0 280,scale=0.5]{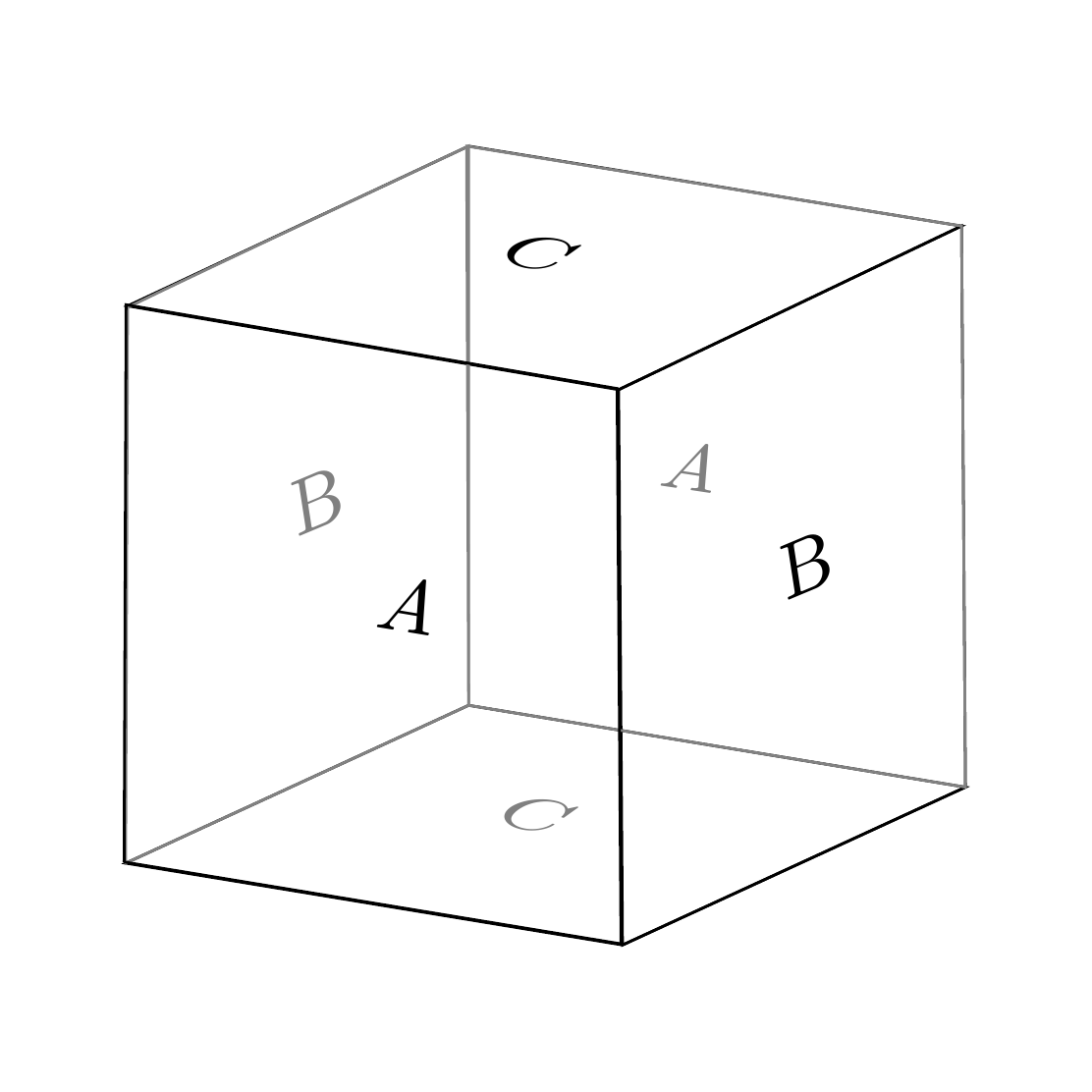}
			\caption{In the same way that a two-dimensional torus can be constructed by gluing the opposite edges of the square, it is perfectly possible to design a three-dimensional torus by an analogous gluing procedure. Despite the practical impossibility of visualizing the result of this construction, it is sufficient to take a three-dimensional cube and to glue its opposite faces. Note that, here (and here only), the letters in the figure do not refer to any of the operators mentioned in these notes: these letters serve only as indexes that highlight the faces that, \textquotedblleft two by two\textquotedblright , need to be glued together.}
			\label{cubo-toro}
		\end{figure}
			\item for the second lattice (which is not infinite but periodic), the degeneracy of the ground state is likely to be greater because this case is analogous to that of the TC, since any cubic lattice with periodic boundary conditions in all three directions can be perfectly identified, by construction, as the cubic discretization of a three-dimensional torus $ \mathcal{T} _{3} $ (see Figure \ref{cubo-toro}).
		\end{itemize}
		
		Indeed, by remembering that the TC degeneracy is directly related to the existence of non-contractile curves in $ \mathcal{T} _{2} $, if we really want to understand how the ground state degeneracy of a new three-dimensional Toric Code ($ 3 $TC) works, we need to understand the non-contractility that is related to $ \mathcal{T} _{3} $. And certainly one of the things we \emph{could} use for this purpose is the simple fact that the equivalence classes that define $ \pi _{1} \left( \mathcal{T} _{3} \right) $ is equal to three: after all, in the same way that happens in the TC, the non-contractility of the curves which belong to each of these classes \emph{could} lead us to new vacuum states independent of (\ref{vacuo-3d-1}).
			
		But, as we emphasize in this last statement, \textquotedblleft could\textquotedblright . Because, as the action of
		\begin{equation*}
			O^{x} _{\bar{\boldsymbol{\gamma }} ^{\ast }} = \prod _{j \in \bar{\boldsymbol{\gamma }} ^{\ast }} \sigma ^{x} _{j}
		\end{equation*}
		no longer corresponds to vacuum when done along any closed path $ \bar{\boldsymbol{\gamma }} ^{\ast } $, the elements of the fundamental group $ \pi _{1} \left( \mathcal{T} _{3} \right) $ cannot moderate any vacuum state of the $ 3 $TC. Yet, since we are dealing with a three-dimensional torus, there is still a topological aspect that can be explored to assess the ground state degeneracy of this $ 3 $TC: this aspect is the order of the \emph{homology groups} $ \mathcal{H} _{k} \left( \mathcal{T} _{3} \right) $, since each of them measures the amount of $ k $-cycles of $ \mathcal{T} _{3} $ that cannot be considered as $ k $-boundaries, where $ 0 \leqslant k \leqslant 3 $ \cite{jessica}. After all, just as it is not difficult to demonstrate that there are $ n $ classes of non-contractile closed curves that provide structure to the fundamental group of an $ n $-dimensional torus $ \mathcal{T} _{n} $, it is also not difficult to see, for instance, that there are non-contractile tori $ \mathcal{T} _{k} $ embedded in $ \mathcal{T} _{n} $.
			
		Although all this seems to be quite different from what is involved in two-dimensional tori, if we look closely at the non-contractile closed curves that can be defined in $ \mathcal{T} _{2} $, this \textquotedblleft bad impression\textquotedblright \hspace*{0.01cm} begins to undo quickly: as these curves have periodic boundary conditions, they must be seen as the $ 1 $-cycles that are not boundaries of this two-dimensional torus and, therefore, they can be seen as the one-dimensional tori $ \mathcal{T} _{1} $ embedded in $ \mathcal{T} _{2} $. In this way, we can affirm that all these \textquotedblleft new\textquotedblright \hspace*{0.01cm} surfaces $ \mathcal{T} _{k} $ are no more than simple generalizations of curves $ \bar{\boldsymbol{\gamma }} = \mathcal{T} _{1} $ on $ \mathcal{T} _{2} $ \footnote{That is, they are the $ k $-cycles we said above, which are contained in $ \mathcal{T} _{n} $ and cannot be seen as $ k $-boundaries.}, because there is a greater freedom to define them which depends even on the fact that $ n $ is a number greater than two\footnote{Note that $ \mathcal{T} _{2} $ can also be seen, for instance, as a non-contractile toroidal surface that is embedded in itself.}.
			
		By effect of this observation, three other vacuum states become evident in this $ 3 $TC:
		\begin{equation}
			\bigl\vert \xi ^{\left( 1 + d \right) } _{0} \bigr\rangle = O^{ \ x} _{\mathcal{T} _{d}} \bigl\vert \xi ^{\left( 1 \right) } _{0} \bigr\rangle \ , \label{$ 3 $-DC-vacuo-degenerado-toro-z}
		\end{equation}	
		where $ d = 1 , 2 , 3 $ and
		\begin{equation}
			O^{ \ x} _{\mathcal{T} _{d}} = \prod _{j \perp \mathcal{T} _{d}} \sigma ^{x} _{j} \label{perp-operator}
		\end{equation}
		is such that each $ \sigma ^{x} _{j} $ acts on the $ j $-th edge which is perpendicular to the discretization of a two-dimensional torus $ \mathcal{T} _{d} \subset \mathcal{T} _{3} $; this discretization occurs by fixing the faces that are normal to one of the three possible directions $ d $. Of course, as well as with (\ref{vacuo-degenerado}), these are not the only additional vacuum states related to the model: as there are \emph{four} possible combinations
		\begin{equation}
			O^{ \ x} _{\mathcal{T} _{1}} \circ O^{ \ x} _{\mathcal{T} _{2}} \ , \quad O^{ \ x} _{\mathcal{T} _{1}} \circ O^{ \ x} _{\mathcal{T} _{3}} \ , \quad O^{ \ x} _{\mathcal{T} _{2}} \circ O^{ \ x} _{\mathcal{T} _{3}} \quad \textnormal{and} \quad O^{ \ x} _{\mathcal{T} _{1}} \circ O^{ \ x} _{\mathcal{T} _{2}} \circ O^{ \ x} _{\mathcal{T} _{3}}
		\end{equation}
		which can be done by using the operators (\ref{perp-operator}), there are also four new vacuum states which make it very clear that we are dealing with a model whose ground state is \emph{eight-fold degenerated}. This eight-fold degeneracy fully agrees with the fact that the second homology group of $ \mathcal{T} _{3} $ is
		\begin{equation*}
			\mathcal{H} _{2} \left( \mathcal{T} _{3} \right) = \mathds{Z} \oplus \mathds{Z} \oplus \mathds{Z} \ .
		\end{equation*}
		After all, since the first $ \mathcal{H} _{1} \left( \mathcal{T} _{3} \right) $ and the second $ \mathcal{H} _{2} \left( \mathcal{T} _{3} \right) $ homology groups of a three-dimensional torus are equal, by noting the \emph{Hurewicz Theorem} shows us that $ \mathcal{H} _{1} \left( \mathcal{T} _{3} \right) $ can be obtained through an abelianization of a $ \pi _{1} \left( \mathcal{T} _{3} \right) $ \cite{hatcher} that is composed of \emph{eight} homotopy classes, we can use this vision to make the following statement: in the same way that $ \mathcal{H} _{1} \left( \mathcal{T} _{3} \right) $ allows us to identify the three generators that lead to the eight distinct homotopy classes defining $ \pi _{1} \left( \mathcal{T} _{3} \right) $, we can associate each one of the three elements of $ \mathcal{H} _{2} \left( \mathcal{T} _{3} \right) $ with the generators that lead to the eight distinct homotopy classes that define the second homotopy group, $ \pi _{2} \left( \mathcal{T} _{3} \right) $. Thus, as the eight combinations of the \emph{non-contractile closed surfaces} mentioned above, which index the eight vacuum states of the $ 3 $TC, correspond exactly to these eight distinct homotopy classes of $ \pi _{2} \left( \mathcal{T} _{3} \right) $, we can affirm that this three-dimensional Toric Code has topological order.
	
	\section{Final remarks}
	
		According to all that we have just presented, what justifies the topological order in the TC is the fact that this model is defined in a two-dimensional torus $ \mathcal{T} _{2} $. The fact that its vacuum state is not unique, for instance, is associated with the one-to-one relationship that exists among these vacuum states and the combinations that can be made using the generators of the fundamental group of $ \mathcal{T} _{3} $. The fact that the statistics (of some) of their quasiparticles do not correspond to bosons or fermions is related to this two-dimensionality: besides these excitations do not have the same transport freedom which they would have in a three-dimensional manifold, it is worth to emphasize that particle systems on compact surfaces (such as a two-dimensional torus) actually present rational statistics as mentioned in \cite{lerda-anyons}.
		
		Of course, since free particles in a three-dimensional manifold have greater freedom of transit than in a two-dimensional manifold, it was to be expected that all conceivable excitations in the $ 3 $TC would be identified only either bosons or fermions \cite{braibant}. Yet, although a different statistic is non-explicit in this three-dimensional generalization, a topological order is still present: the degeneracy degree of the $ 3 $TC ground state is equal to the number of homotopy classes in $ \pi _{2} \left( \mathcal{T} _{3} \right) $; i.e., we have a degeneracy that, because it is topological, allows us to affirm that the Toric Code also has a topological order in the three-dimensional case.
		
		Lastly, it is worth noting that the statistics of any particle can only be obtained, for example, in a situation where it is possible to exchange particles by keeping the system energy as a constant. And this is what justifies the definition that we give for one of the $ 3 $DC quasiparticles, the \textquotedblleft quasiplaque\textquotedblright , as a combination of four elementary excitations that are identifiable as the same magnetic quasiparticles $ m $ of the TC: if this definition were not made in this way, the energy of the system would increase in a manner similar to that of a quark when removed from another. In any case, this energy increase in our model follows the same perimeter law that, for example, has already been observed in models similar to the two-dimensional TC, as in the lattice gauge theories mentioned in Ref. \cite{alcaraz}, and in the models that we are developing by using one generalization of the QDM, which will be the central theme of our next paper, suggesting that there is a connection among all these models.
		
	\section*{Acknowledgements}
	
		This work has been supported by CAPES (ProEx) and CNPq (grant 162117/2015-9). We thank L. D. Borsari, D. L. Gon\c{c}alves, J. P. Ibieta Jimenez, J. Lorca Espiro, R. Figueiredo, D. V. Tausk and P. Teotonio Sobrinho for some discussions on subjects concerning this project, U. A. Maciel Neto and D. A. Soares for some suggestions on writing which helped to enrich this text, and to good person who published in the Ref. \cite{website}, by using the codname \textquotedblleft AJN\textquotedblright , the latex code we used to generate the two-dimensional torus in Figure \ref{torus} that we edit to add two non-contractile curves.


\begin{thebibliography}{50}
		\bibitem{kitaev-conference} A. Kitaev, Quantum error correction with imperfect gates, in {\it Proceedings of the 3rd International Conference of Quantum Communication and Measurement}, pp. 181-188, Ed. O. Hirota, A. S. Holevo and C. M. Caves (Plenum Press, New York, 1997).
		\bibitem{fragility} Z. Nussinov and G. Ortiz, Autocorrelations and thermal fragility of anyonic loops in topologically quantum ordered systems, {\it Phys. Rev. B} \textbf{77} (2008) 064302.
		\bibitem{kitaev-1} A. Yu. Kitaev, Fault-tolerant quantum computation by anyons, {\it Annals Phys.} \textbf{303} (2003) pp. 2-30.
		\bibitem{pachos} J. K. Pachos, {\it Introduction to Topological Quantum Computation} (Cambridge University Press, New York, 2012).
		\bibitem{pervin} W. J. Pervin, {\it Foundations of General Topology}, (Academic Press, New York, 1964).
		\bibitem{bellac} M. L. Bellac, {\it A Short Introduction to Quantum Computation and Quantum Computation} (Cambridge University Press, Cambridge, 2006).
		\bibitem{mf} M. F. Araujo de Resende, {\it Quantiza\c{c}\~{a}o da part\'{\i}cula não relativ\'{\i}stica em espa\c{c}os curvos como superf\'{\i}cies do $ \mathds{R} ^{\textnormal{n}} $} (Disserta\c{c}\~{a}o de Mestrado IFUSP, S\~{a}o Paulo, 2011).
		\bibitem{ballen} L. E. Ballentine, {\it Quantum Mechanics: A Modern Development} (World Scientific, Singapore, 2000).
		\bibitem{kitaev-math} A. Kitaev, Topological quantum codes and anyons, {\it Proceedings of Symposia in Applied Mathematics} \textbf{58} (2002) pp. 267-272.
		\bibitem{elon} E. L. Lima, {\it \'{A}lgebra Linear} (IMPA, Rio de Janeiro, 2008).
		\bibitem{arfken} G. Arfken, {\it Mathematical Methods for Physics, Third Edition}, (Academic Press Inc., San Diego, 1985).
		\bibitem{cirac} F. Verstraete, M. M. Wolf, D. Perez-Garc\'{\i}a and J. I. Cirac, Criticality, the Area Law, and the Computational Power of Projected Entangled Pair States, {\it Phys. Rev. Lett.} \textbf{96} (1985) 220601.
		\bibitem{bianconi} R. Bianconi, {\it Geometria e Desenho Geom\'{e}trico I} (Notas de Aula e Exerc\'{\i}cios IME-USP, S\~{a}o Paulo, 2001).
		\bibitem{hatcher} A. Hatcher, {\it Algebraic Topology} (Cambridge University Press, New York, 2002).
		\bibitem{castelnovo} C. Castelnovo, S. Trebst and M. Troyer, Topological Order and Quantum Criticality, in {\it Understanding Quantum Phase Transitions}, Ed. L. D. Carr (CRC Press Taylor \& Francis Group, 2011).
		\bibitem{itzykson} C. Itzykson and J. B. Zuber, {\it Quantum Field Theory} (McGraw-Hill Inc., New York, 1980).
		\bibitem{aha-bohm} Y. Aharonov and D. Bohm, Significance of Electromagnetic Potentials in the Quantum Theory, {\it Phys. Rev.} \textbf{115} (1959) pp. 485-491.
		\bibitem{lerda-anyons} A. Lerda, {\it Anyons: Quantum Mechanics of Particles with Fractional Statistics} (Springer-Verlag, Berlin Heidelberg, 1992).
		\bibitem{jessica} J. C. R. R. Costa and M. G. Carreira Andrade, Algumas considera\c{c}\~{o}es sobre homotopia e homologia, {\it CQD - Revista Eletr\^{o}nica Paulista de Matem\'{a}tica} \textbf{2} (2013) pp. 18-31.
		\bibitem{braibant} S. Braibant, G. Giacomelli and M. Spurio, {\it Particles and Fundamental Interactions: An Introduction to Particle Physics} (Springer Netherlands, Dordrecht, 2012).
		\bibitem{alcaraz} F. C. Alcaraz, {\it Uma introdu\c{c}\~{a}o em teorias de calibre na rede} (Notas de aula, Escola de F\'{\i}sica Jorge Andr\'{e} Swieca, 1982).
		\bibitem{website} https://tex.stackexchange.com/questions/304585/customizing-torus-diagram-in-pgfplots/304596$ \# $304596
	\end{thebibliography}
\end{document}